\documentclass[aps,prd,notitlepage,twoside,onecolumn,preprintnumbers,showpacs,showkeys,byrevtex,floatfix,nofootinbib,amsmath,amssymb,superscriptaddress]{revtex4-1}

\usepackage{amsmath,amssymb,amsthm,array,booktabs,dsfont,graphicx,mathtools,multirow,rotating,stmaryrd,tensor,verbatim}
\usepackage[bookmarksnumbered,linktocpage,pdfstartview=FitH]{hyperref}
\usepackage[all]{hypcap}
\usepackage[margin=2cm]{geometry}

\graphicspath{{figures/}}

\newcolumntype{M}[1]{>{$}{#1}<{$}}
\newcolumntype{C}[1]{>{\centering}m{#1}}

\newcommand{\rep}[1]{\ensuremath{\mathbf{#1}}}

\newcolumntype{B}[1]{>{\mathbf\bgroup}{#1}<{\egroup}}
\newcolumntype{K}{>{\lvert}{c}<{\rangle}}

\newtheorem{theorem}{Theorem}
\newtheorem*{theorem*}{Theorem}

\DeclareMathOperator{\Aut}{Aut}
\DeclareMathOperator{\Str}{Str}

\DeclareMathOperator{\Tr}{Tr} 
\DeclareMathOperator{\Det}{Det}

\DeclareMathOperator{\Iso}{Iso}
\DeclareMathOperator{\Hom}{Hom}
\DeclareMathOperator{\SO}{SO}
\DeclareMathOperator{\Usp}{Usp}
\DeclareMathOperator{\SL}{SL}
\DeclareMathOperator{\SU}{SU}
\DeclareMathOperator{\Sp}{Sp}

\DeclareMathOperator{\U}{U}

\newcommand{\pmtwo}[4]{\begin{pmatrix}#1 & #2 \\ #3 & #4 \end{pmatrix}}
\newcommand{\pmthree}[9]{\begin{pmatrix}#1 & #2&#3 \\ #4 & #5&#6 \\#7 & #8&#9  \end{pmatrix}}
\newcommand{\be}{\begin{equation}}
\newcommand{\ee}{\end{equation}}
\newcommand{\bea}{\begin{eqnarray}}
\newcommand{\eea}{\end{eqnarray}}

\newcommand{\half}{\tfrac{1}{2}}

\newcommand{\J}{\mathfrak{J}}
\newcommand{\JOs}{\mathfrak{J}^{\mathds{O}^s}_{3}}
\newcommand{\JA}{\mathfrak{J}^{\mathds{A}}_{3}}

\newcommand{\JO}{\mathfrak{J}^{\mathds{O}}_{3}}
\newcommand{\JH}{\mathfrak{J}^{\mathds{H}}_{3}}
\newcommand{\JC}{\mathfrak{J}^{\mathds{C}}_{3}}
\newcommand{\JR}{\mathfrak{J}^{\mathds{R}}_{3}}

\newcommand{\alg}{\mathds{A}}

\newcommand{\F}{\mathds{F}}
\newcommand{\R}{\mathds{R}}
\newcommand{\C}{\mathds{C}}
\newcommand{\Q}{\mathds{H}}
\newcommand{\Z}{\mathds{Z}}
\newcommand{\Oct}{\mathds{O}}
\newcommand{\sO}{\mathds{O}^s}

\newcommand{\FTS}{\mathfrak{F}}
\newcommand{\FOs}{\mathfrak{F}^{\mathds{O}^s}}
\newcommand{\FO}{\mathfrak{F}^\Oct}
\newcommand{\FH}{\mathfrak{F}^\Q}
\newcommand{\FC}{\mathfrak{F}^\C}
\newcommand{\FTSR}{\mathfrak{F}^\R}
\newcommand{\FA}{\mathfrak{F}^\alg}

\newcommand{\Fnt}{\mathfrak{F}^{2,n}}
\newcommand{\Fnf}{\mathfrak{F}^{6,n}}
\newcommand{\Rstr}[1]{\mathfrak{Str}_0(#1)}
\newcommand{\AutF}[1]{\Aut(#1)}

\newcommand{\Jnt}{\mathfrak{J}_{1,n-1}}
\newcommand{\Jnf}{\mathfrak{J}_{5,n-1}}
\newcommand{\Ne}{\mathcal{N}=8}

\begin{document}

\title{Small Orbits}
\author{L. Borsten}
\email[]{leron.borsten@imperial.ac.uk}
\affiliation{INFN Sezione di Torino \&  Dipartimento di Fisica Teorica,
Universit\`a di Torino  Via Pietro Giuria 1, 10125 Torino, Italy}
\author{M. J. Duff}
\email[]{m.duff@imperial.ac.uk}
\affiliation{Theoretical Physics, Blackett Laboratory, Imperial College London, London SW7 2AZ, United Kingdom}
\author{S. Ferrara}
\email[]{sergio.ferrara@cern.ch}
\affiliation{Physics Department, Theory Unit, CERN, CH -1211, Geneva 23, Switzerland}
\affiliation{INFN - Laboratori Nazionali di Frascati, Via Enrico Fermi 40, I-00044 Frascati, Italy}
\affiliation{Department of Physics and Astronomy, University of California, Los Angeles, CA 90095-1547,USA}
\author{A. Marrani}
\email[]{alessio.marrani@cern.ch}
\affiliation{Physics Department, Theory Unit, CERN, CH -1211, Geneva 23, Switzerland}
\setcounter{affil}{3}
\author{W. Rubens}
\email[]{william.rubens06@imperial.ac.uk}
\affiliation{Theoretical Physics, Blackett Laboratory, Imperial College London, London SW7 2AZ, United Kingdom}

\date{\today}

\begin{abstract}
We study both the ``large'' and ``small'' U-duality charge orbits of
extremal black holes appearing in $D=5$ and $D=4$ Maxwell-Einstein
supergravity theories with symmetric scalar manifolds. We exploit a
formalism based on cubic Jordan algebras and their associated
Freudenthal triple systems, in order to derive the minimal charge
representatives, their
stabilizers and the associated ``moduli spaces''. After recalling $\mathcal{N%
}=8$ maximal supergravity, we consider $\mathcal{N}=2$ and
$\mathcal{N}=4$
theories coupled to an arbitrary number of vector multiplets, as well as $%
\mathcal{N}=2$ magic, $STU$, $ST^{2}$ and $T^{3}$ models. While the
$STU$ model may be considered as part of the general $\mathcal{N}=2$
sequence, albeit with an additional triality symmetry, the $ST^{2}$
and $T^{3}$ models demand a separate treatment, since their
representative Jordan algebras are Euclidean or only admit non-zero
elements of rank 3, respectively. Finally,
we also consider \textit{minimally coupled} $\mathcal{N}=2$, matter coupled $%
\mathcal{N}=3$, and ``pure'' $\mathcal{N}=5$ theories.
\end{abstract}



\preprint{Imperial/TP/2011/mjd/2, CERN-PH-TH/2011-162}

\maketitle


\tableofcontents
\section{Introduction}
\subsection{Background}
A concerted effort has been made to understand the physically
distinct black hole (BH) solutions appearing in various
4-dimensional supergravity theories. The extremal  solutions
typically carry electromagnetic charges transforming linearly under
$G_4$, the $D=4$ U-duality group\footnote{We work in the classical
regime for which the electromagnetic charges are real valued. Here
U-duality $G_4$ is referred to as the ``continuous" symmetries of
\cite{Cremmer:1979up}. Their discrete versions are the
non-perturbative U-duality string theory symmetries described in
\cite{Hull:1994ys}.}. BHs with charges lying in different orbits of
$G_4$ therefore correspond to distinct solutions. Moreover, thanks
to the attractor mechanism
\cite{Ferrara:1995ih,Strominger:1996kf,Ferrara:1996dd,Ferrara:1996um,Ferrara:1997tw}
the entropy of the extremal BH solutions loses all memory of the
scalars at infinity and is a function of only the charges.
Consequently, the Bekenstein-Hawking
\cite{Bekenstein:1973ur,Bardeen:1973gs} entropy is given by a
U-duality invariant quartic in the electromagnetic charges. Hence,
the classification of the U-duality charge  orbits captures many
significant features of the possible BH solutions, which in turn
have provided a range of important string or M-theoretic insights.

 We focus on those theories in which the scalars live in a symmetric  coset $G_4/H_4$. The orbits of the 4-dimensional $\mathcal{N}=8$ \cite{Cremmer:1979up} and the exceptional octonionic ``magic'' $\mathcal{N}=2$ \cite{Gunaydin:1983rk} supergravities  were obtained in \cite{Ferrara:1997uz} for both ``large'' and ``small'' BHs, which have non-vanishing or vanishing classical entropy, respectively. The large orbits of the $\mathcal{N}=2$ Maxwell-Einstein supergravities coupled to $n_V$ vector multiplets, which also include the three non-exceptional magic examples, were analysed in \cite{Ferrara:1997uz, Bellucci:2006xz}. The small orbits of the $STU$ model \cite{Cvetic:1995uj,Duff:1995sm,Cvetic:1995bj,Cvetic:1996zq,Behrndt:1996hu,Bellucci:2007zi, Bellucci:2008sv}, which exhibits a discrete triality, exchanging the roles of $S$, $T$ and $U$, over and above the continuous U-duality group,  were found in \cite{Borsten:2009yb}. Meanwhile, for the infinite sequence of $\mathcal{N}=4,2$ theories coupled to $n_V$ vector multiplets the U-duality invariant charge constraints defining the distinct orbits and their supersymmetry preserving properties, for both large and small cases,  were obtained in \cite{Ferrara:1997ci,Cerchiai:2009pi}, and further discussed in \cite{Andrianopoli:2010bj,Ceresole:2010nm}.

In the present work, we aim at essentially completing this story in
$D=4$. In particular, we obtain the small orbits for the  $\mathcal{N}=2$ $\R, \C, \Q$ magic supergravities,  $\mathcal{N}=2,4$ supergravity coupled to an arbitrary number of vector multiplets including  the special cases of the $STU$, $ST^2$ and $T^3$ models, as well as the \textit{minimally coupled} $\mathcal{N}=2$, matter coupled
$\mathcal{N}=3$, and ``pure" $\mathcal{N}=5$ theories.

 We begin by repeating  the $\mathcal{N}=8$ theory as it
provides an instructive example, setting the stage for all the other
cases. We then study both the ``large'' and ``small'' U-duality BH
charge orbits of the $D=4$, $\mathcal{N}=4$ and $\mathcal{N}=2$
Maxwell-Einstein supergravity theories coupled to an arbitrary
number $n_V$ of vector multiplets, including the magic theories. The
$\mathcal{N}=2$ $STU$ model is retreated as part of the generic
sequence ($n_V=3$), revealing additional subtleties which were
previously obscured by  the triality symmetry. Its degeneration into
the $ST^2$ and $T^3$ models is also treated.  A formalism  based on
 cubic Jordan algebras and their associated Freudenthal
triple systems (FTS) is used to derive the minimal charge orbit
representatives, their stabilizers and the associated ``moduli
spaces'' of attractor solutions. In particular, we make use of
\cite{Borsten:2011nq} and \cite{Krutelevich:2004,Shukuzawa:2006}.
While the $STU$ model may be considered as  part of the general
$\mathcal{N}=2$ sequence, albeit with an additional triality
symmetry, the $ST^2$ and $T^3$ models demand a separate treatment. This is due to their representative Jordan algebras being, in some sense, degenerate:  the $ST^2$ Jordan algebra is  Euclidean, as opposed to the Lorentzian nature of the general sequence, while the $T^3$ Jordan algebra only contains non-zero elements of rank 3. Finally, in 
\autoref{N=2-d=4-quadratic}, \autoref{N=3,d=4} and \autoref{N=5,d=4}, we
respectively include the analogous treatment of the
\textit{minimally coupled} $\mathcal{N}=2$, matter coupled
$\mathcal{N}=3$, and ``pure" $\mathcal{N}=5$ theories, which cannot
all be uplifted to $D=5$ space-time dimensions.

 Physically speaking, the FTS makes the symmetries of the parent $D=5$ theory manifest. This allows us to make extensive use of the orbits and their minimal charge representatives of the $D=5$ theories, which are simpler to derive and already appeared in the literature. In particular, we exploit the analysis of \cite{Ferrara:1997uz,Ferrara:2006xx,Ceresole:2007rq,Cerchiai:2009pi,Cerchiai:2010xv,Ceresole:2010nm}. Note, one may also use the \emph{integral} FTS to address the orbit classification of the discrete stringy U-duality groups \cite{Hull:1994ys}, as was done for the maximally supersymmetric $D=6,5,4$ theories in \cite{Borsten:2009zy,Borsten:2010aa}.  Moreover, for $D=4, \mathcal{N}=8$ it has recently been observed that some of the orbits
of $E_{7(7)}(\Z)$ should play an important role in counting
microstates of this theory \cite{Bianchi:2009wj,Bianchi:2009mj}. The
importance of discrete invariants and orbits to the dyon spectrum of
string theory has been the subject  of much investigation
\cite{Dabholkar:2007vk,Sen:2007qy,Banerjee:2007sr,Banerjee:2008pu,Banerjee:2008ri,Sen:2008ta,Sen:2008sp,Bianchi:2009mj}.

\subsection{Summary}

We summarise the key results here. For each of the theories
considered (aside from the $\mathcal{N}=2$
minimally coupled, $\mathcal{N}=3$ and $\mathcal{N}=5$ theories), the electromagnetic BH charges may be regarded as
elements of a Freudenthal triple system \be
\mathfrak{F}(\J_3):=\R\oplus\R\oplus\J_3\oplus\J_3, \ee
 defined over a cubic Jordan algebra $\J_3$. The electric (magnetic) BH (black string - BS -) charges of the  parent $D=5$ theory may be regarded as elements of $\J_3$. The FTS comes equipped with three maps: (i) a bilinear antisymmetric form $\{\bullet, \bullet\}:\FTS\times\FTS\rightarrow \R$, which encodes the symplectic structure of the charge representations (see for example \cite{Andrianopoli:1996ve}, and Refs. therein); (ii) a quartic norm $\Delta:\FTS\rightarrow \R$; (iii) a triple product $T:\FTS\times\FTS\times\FTS\rightarrow\FTS$. A brief summary may be found in \autoref{sec:FTS}. Full details can be found in \cite{Borsten:2011nq} and  Refs. therein. The automorphism group $\Aut(\FTS)\cong \text{Conf}\left( \frak{J}_{3}\right)$ is the set of invertible $\R$-linear transformations preserving the quartic norm and bilinear form. It coincides with the $D=4$ U-duality group: $\Aut(\FTS)=G_4$. Hence, the unique quartic $G_4$-invariant, denoted $I_4$, is given by $\Delta$. The Bekenstein-Hawking entropy
 therefore reads
\be \label{BH-entropy-D=4}
S_{\text{BH}}=\pi\sqrt{|\Delta|}=\pi\sqrt{|I_4|}. \ee

Let us briefly review some of the analogous features of  cubic
Jordan algebras and the BHs (BSs) in $D=5$, which we will make
extensive use of throughout. A cubic Jordan algebra $\J_3$ is a
vector space equipped with an admissible cubic norm $N:
\J_3\rightarrow\R$ and an element $c\in\J_3$, referred to as a
\emph{base point}, satisfying $N(c)=1$. The cubic norm defines the
Jordan product, $-\circ-:\J_3\times\J_3\rightarrow\J_3$, satisfying,
\be X^2\circ(X\circ Y)=X\circ(X^2\circ Y), \qquad \forall X, Y \in
\J_3. \ee A brief summary may be found in \autoref{sec:FTS}. Full details can be found in \cite{Borsten:2011nq} and  Refs. therein. For each of the theories
considered in the present investigation (but the $\mathcal{N}=2$
minimally coupled, $\mathcal{N}=3$ and $\mathcal{N}=5$ theories),
the electromagnetic BH charges may be regarded as elements of some
cubic Jordan algebra $\J_3$. The automorphism group $\Aut(\J_3)$ is
the set of invertible $\R$-linear transformations preserving the
Jordan product. The reduced structure group $\Str_0(\J_3)$ is the
set of invertible $\R$-linear transformations preserving the cubic
norm $N$ \cite{Borsten:2011nq}. $\Str_0(\J_3)$ is the $D=5$ U-duality
group, $ \Str_0(\J_3)=G_5$. Hence, the unique cubic $G_5$-invariant,
denoted $I_3$, is given by $N$. The Bekenstein-Hawking BH (BS)
entropy is therefore \be S_{\text{BH}}=\pi\sqrt{|N|}.
\label{entropy-D=5} \ee

The models we consider  are itemized here:
\begin{itemize}
\item $\mathcal{N}=8$: $28+28$ electric/magnetic BH charges belong to $\FOs:=\FTS(\JOs)$, where $\JOs$ is the cubic Jordan algebra of $3\times 3$ Hermitian matrices defined over the split-octonions. The 56 charges transform linearly as the fundamental $\rep{56}$ of   $\Aut(\FOs)=E_{7(7)}\cong \text{Conf}\left( \frak{J}_{3}^{\mathds{O}_{s}}\right)$, the maximally non-compact (split) real form of $E_7(\C)$. The scalar manifold is given
by (apart from discrete factors, see \textit{e.g.}
\cite{Yokota:2009}) \be \frac{E_{7(7)}}{\SU(8)}. \ee
\item Magic $\mathcal{N}=2$ theories: Given by $\mathcal{N}=2$ supergravity coupled to  $(3+3\dim \alg)$ vector multiplets, where $\alg=\R, \C, \Q, \Oct$. The $(4+3\dim \alg)+(4+3\dim \alg)$ electric/magnetic BH charges belong to $\FA:=\FTS(\JA)$, where $\JA$ is the cubic Jordan algebra of $3\times 3$ Hermitian matrices defined over one of the four division algebras $\alg=\R, \C, \Q, \Oct$. The $(8+6\dim \alg)$ charges transform linearly as the threefold antisymmetric traceless tensor $\rep{14}'$, the threefold antisymmetric self-dual tensor $\rep{20}$, the chiral spinor $\rep{32}$ and the fundamental $\rep{56}$ of   $\Aut(\FA)\cong \text{Conf}\left( \frak{J}_{3}^{\mathds{A}}\right)=\Sp(6, \R), \SU(3,3), \SO^\star(12), E_{7(-25)}$ for $\alg=\R, \C, \Q, \Oct$, respectively. The scalar manifolds are given
by (apart from discrete factors, see \textit{e.g.}
\cite{Yokota:2009})  \be \frac{\Sp(6, \R)}{\U(3)},\quad\frac{\SU(3,
3)}{\U(1)\times
\SU(3)\times\SU(3)},\quad\frac{\SO^\star(12)}{\U(6)},\quad\frac{E_{7(-25)}}{\U(1)\times
E_{6{(-78)}}}. \ee
\item $\mathcal{N}=4$ supergravity ($6$ graviphotons) coupled to $n=n_V$ vector multiplets:
the $(n_V+6)+(n_V+6)$ electric/magnetic BH charges belong to
$\Fnf:=\FTS(\Jnf)$, where $\frak{J}_{5,n-1}\cong \mathds{R}\oplus
\mathbf{\Gamma }_{5,n-1}$ is the cubic Jordan algebra of
pseudo-Euclidean spin factors \cite{McCrimmon:2004} (see also
\cite{Borsten:2011nq}). In general, $\mathbf{\Gamma }_{m,n}$ is a Jordan algebra with a quadratic form of pseudo-Euclidean
signature $\left( m,n\right) $, \textit{i.e.} the Clifford algebra
of $O\left( m,n\right) $ \cite{Jordan:1933vh}. The $2(n_V+6)$
charges transform linearly as the $(\mathbf{2},\mathbf{6+n_V})$ of
$\Aut(\Fnf)\cong \text{Conf}\left( \frak{J}_{5,n-1}\right)
=\SL(2,\R)\times\SO(6, n_V)$. The scalar manifolds are given by the
infinite sequence of globally symmetric Riemannian manifolds \be
\frac{\SL\left( 2,\mathds{R}\right) }{\SO\left( 2\right) }\times \frac{%
\SO\left( 6,n_{V}\right) }{\SO\left( 6\right) \times \SO\left( n_{V}\right) }%
,~n_{V}\geqslant 0. \ee

\item $\mathcal{N}=2$ supergravity ($1$ graviphoton) coupled to $n_V$ vector multiplets: the $(n_V+1)+(n_V+1)$ electric/magnetic BH charges belong to $\Fnt:=\FTS(\Jnt)$, where $\frak{J}_{1,n-1}\cong \mathds{R}\oplus \mathbf{\Gamma }_{1,n-1}$ is the cubic Jordan algebra of Lorentzian spin factors \cite{McCrimmon:2004} (see also \cite{Borsten:2011nq}), and $n=n_V-1$. The $2(n_V+1)$ charges transform linearly as the $(\mathbf{2},\mathbf{1+n_V})$ of $\Aut(\Fnt)\cong \text{Conf}\left( \frak{J}_{1,n-1}\right)=\SL(2,\R)\times\SO(2, n)$.
The scalar manifolds are given by the
infinite sequence of globally symmetric special K\"{a}hler manifolds\be \frac{\SL\left( 2,\mathds{R}\right) }{\SO\left( 2\right) }\times \frac{%
\SO\left( 2,n_{V}-1\right) }{\SO\left( 2\right) \times \SO\left(
n_{V}-1\right) },~n_{V}\geqslant 2. \label{N=2-JS} \ee

\item $\mathcal{N}=2$ $STU$ model: it is nothing but $n_V=3$ element of the Jordan symmetric sequence \eqref{N=2-JS},
but we single it out for two reasons. First, over and above the
continuous U-duality group it has a discrete \textit{triality
symmetry} which swaps the roles of the three complex moduli $S,T,U$
\cite{Duff:1995sm}, and is manifested in the structure of the
duality orbits. Second, it may be considered as the common sector of
all $D=4$ Maxwell-Einstein supergravity theories with a rank-3
symmetric vector multiplets' scalar manifold and related to Jordan
algebras (which we will dub \textit{``symmetric''} supergravities).
Furthermore, it also provides a link to the degenerate cases
described below. The $4+4$ electric/magnetic BH charges belong to
$\FTS_{STU}:=\FTS(\J_{STU})$, where $\J_{STU}=\R\oplus\R\oplus\R$ is
isomorphic to the  Lorentzian spin factor $\J_{1,1}$
\cite{McCrimmon:2004,Borsten:2011nq}. The 8 charges transform
linearly as the $(\rep{2, 2, 2})$ of $\Aut(\FTS_{STU})\cong
\text{Conf}\left(
\frak{J}_{STU}\right)=\SL(2,\R)\times\SL(2,\R)\times\SL(2,\R)$. This
symmetry is made manifest by organising the charges into a $2\times
2\times 2$ hypermatrix $a_{ABC}$, where $A,B,C=0,1$, transforming
under $\SL_A(2,\R)\times\SL_B(2,\R)\times\SL_C(2,\R)$
\cite{Duff:2006uz}. The scalar manifold is given by \be \left[
\frac{\SL\left( 2,\mathds{R}\right) }{\SO\left( 2\right) }\right]
^{3}. \label{STU-sc} \ee It is worth noting that, by using
U-duality, the charge vectors of the \textit{symmetric} supergravity
theories described above may be reduced to a subsector living in
$\FTS_{STU}$. Hence, the $STU$ charges are common to all the above
theories which, indeed, may all be consistently truncated to the
$STU$ model. Moreover, the special K\"{a}hler geometry
characterising the completely factorised rank-3 symmetric manifold
(\ref{STU-sc}) is defined by the triality-symmetric prepotential
\be\label{eq:stuprepot} F=STU. \ee See, for example,
\cite{Cremmer:1984hj,deWit:1984px,Strominger:1990pd,Ferrara:1995ih}
for the details of special geometry. By identifying  $T=U$ and
$S=T=U$ in \eqref{eq:stuprepot} we obtain the $ST^2$ and $T^3$
models, respectively (see \textit{e.g.} \cite{Bellucci:2007zi} for
the consistent exploitation of such a degeneration/reduction
procedure). In this sense, the $STU$ model is the linchpin of all
the theories considered here.

\item $\mathcal{N}=2$ $ST^2$ model: coupled to two vector multiplets.
The $3+3$ electric/magnetic BH charges belong to
$\FTS_{ST^2}:=\FTS(\J_{ST^2})$, where $\J_{ST^2}=\R\oplus\R$ is
isomorphic to the Euclidean spin factor $\J_{1}$
\cite{McCrimmon:2004,Borsten:2011nq}. The 6 charges transform
linearly as the $(\rep{2, 3})$ of
$\Aut(\FTS_{ST^2})=\SL(2,\R)\times\SL(2, \R)$. This symmetry is made
manifest by organising the charges into a partially symmetrised
hypermatrix $a_{A(B_1B_2)}$, where $A,B_1, B_2=0,1$, transforming
under $\SL_A(2,\R)\times\SL_B(2,\R)$ \cite{Bellucci:2007zi}. The
scalar manifold is given by \be \left[ \frac{\SL\left(
2,\mathds{R}\right) }{\SO\left( 2\right) }\right] ^{2}. \ee

\item $\mathcal{N}=2$ $T^3$ model:  this is a \textit{non-generic} irreducible model, coupled to a single vector multiplet. May be obtained as a circle compactification of minimal supergravity in five dimensions. The $2+2$ electric/magnetic BH charges belong to $\FTS_{T^3}:=\FTS(\J_{T^3})$, where $\J_{T^3}=\R$. The 4 charges transform linearly as the $\rep{4}$ (spin $s=3/2$) of $\Aut(\FTS_{T^3})\cong \text{Conf}\left( \frak{J}_{T^{3}}\right)=\SL(2,\R)$. This symmetry is made manifest by organising the charges into a totally symmetrised hypermatrix $a_{(A_1A_2A_3)}$, where $A_1, A_2, A_3=0,1$, transforming under $\SL_A(2,\R)$
\cite{Bellucci:2007zi} (see also e.g. \cite{Borsten:2008wd}, as well
as the recent discussion in \cite{Levay:2010ua}). The scalar
manifold is given by the special K\"{a}hler manifold (with scalar
curvature $R=-2/3$ \cite{Cremmer:1984hc}) \be
\frac{\SL(2,\R)}{\SO(2)}. \ee
\end{itemize}

In all aforementioned cases, excluding the $T^3$ model, the charge
orbits are split into four classes first identified in
\cite{Ferrara:1997uz}. There are three small classes with vanishing
Bekenstein-Hawking entropy: doubly critical, critical and
light-like. There is one large class with non-zero
Bekenstein-Hawking entropy, which actually is a one-parameter
($I_4$) family of orbits. The $T^3$ model is the exception in that
the doubly critical and critical classes collapse into a single
orbit. This is precisely due to the fact that the underlying cubic
Jordan algebra $\J_{T^3}$ only admits non-zero elements of rank 3, as
opposed to the other examples, which  all possess elements of rank 1,2 and 3 (including the
$ST^2$ model). From a physical perspective, this is equivalent to
the fact that there is only one gauge potential (namely, only one
Abelian vector multiplet) outside the gravity multiplet to support
both the doubly critical and critical orbits.

These four classes are coded in the ``rank'' of the FTS element:
ranks 1, 2, 3 and 4 imply doubly critical, critical, light-like and
large, respectively. For the $\Ne$ (maximal supersymmetry) theory
the ranks are sufficient to capture all the orbit details,
\textit{i.e.} there is precisely one orbit per rank. The only
subtlety is that the large BHs are supported by a 1/8-BPS or a
non-BPS orbit, according as $I_4>0$ or $I_4<0$, respectively
\cite{Ferrara:1997uz}. For theories of gravity with non-maximal
local supersymmetry, this identification between rank and orbit
generally becomes more subtle: while rank 1 (doubly critical)
elements lie in a single orbit, higher ranks split into two or more
orbits. Moreover, BHs with $I_4>0$ may also be non-BPS; in contrast,
all BHs with $I_4<0$ are non-BPS. In every case, there is only one
$I_4<0$ orbit.

\medskip

We summarise the key features of this orbit splitting here, while
laying out the organisation of the letter.

First, let us mention that the technical aspects of Jordan algebras,
the FTS and the proofs of the associated theorems used here  may be
found in \cite{Borsten:2011nq} and in Refs. therein.  We begin in
\autoref{sec:D5} with a summary of the $D=5$ parent theories: their
Jordan algebras, minimal charge orbit representatives, cosets and
\textit{moduli spaces}. This lays the foundations for the $D=4$
analysis. In \autoref{sec:D4} the details of $D=4$ minimal charge
orbit representatives, cosets and \textit{moduli spaces} are
presented for each of the aforementioned theories. The
$\mathcal{N}=8$ treatment, while having been well understood for
sometime now \cite{Ferrara:1997uz,Borsten:2010aa}, is given first as
the  simplest example (only one orbit per rank of FTS element), with
ranks 1, 2, 3 corresponding to 1/2-, 1/4- and 1/8-BPS states,
respectively. As mentioned, the unique subtlety is that the rank 4
large orbit is 1/8-BPS or non-BPS orbit according as $I_4>0$ or
$I_4<0$. The orbits and their representatives are given in
\autoref{tab:D4N8} and \autoref{thm:N8}, respectively. Also, notice
that the supersymmetry BPS-preserving features are not sufficient to
uniquely characterise the charge orbits; indeed, there are two
1/8-BPS orbits, one large (rank 4) and one small lightlike (rank 3).
All subsequent treatments may be seen as a fine-graining of the
treatment of $\mathcal{N}=8$ orbits. Only the rank 1 (doubly
critical) and the rank 4  $(I_4<0)$ cases do not split, remaining as
a single 1/2-BPS and non-BPS orbit, respectively, for all
non-maximally supersymmetric theories. The next simplest cases are
the magic $\mathcal{N}=2$ supergravities. Here the rank 2, 3 and 4
$(I_4>0)$ orbits split into one 1/2-BPS and non-BPS orbit each. The
non-BPS large $(I_4>0)$ orbit has vanishing central charge at the
unique BH event horizon. The orbits and their representatives are
given in \autoref{tab:Mcharges} and \autoref{thm:Shuk},
respectively. The exceptional octonionic case is given as a detailed
example in \autoref{sec:orbitsmagicFTS}, which thus provides an
alternative derivation of the result obtained in
\cite{Ferrara:1997uz}. Next, comes $\mathcal{N}=4$ Maxwell-Einstein
supergravity. The major difference is that the corresponding FTS is
\textit{reducible}. As a consequence, as proved in
\cite{Borsten:2011nq}, an extra rank 2 orbit is introduced, making a
total of three: 1/2-BPS, 1/4-BPS and non-BPS. Rank 3 has one 1/4-BPS
and one non-BPS, as does rank 4 $(I_4>0)$. The orbits and their
representatives are given in \autoref{tab:n4redorbits} and
\autoref{thm:N4}, respectively. Finally, we consider $\mathcal{N}=2$
Maxwell-Einstein supergravity based on the Jordan symmetric sequence
(\ref{N=2-JS}), which has the most intricate orbit structure.
However, it may be derived directly from the $\mathcal{N}=4$ case by
splitting each 1/4-BPS orbit into one 1/2-BPS and one non-BPS (with
vanishing central charge at the horizon); see
\autoref{sec:N4N2}. We conclude with the ``degenerate'' cases of
$ST^2$ (non-generic reducible) and $T^3$ (non-generic irreducible)
$\mathcal{N}=2$, $D=4$ supergravity models in \autoref{sec:degen}.

Finally, we consider the remaining D=4 theories with symmetric
scalar manifolds, which cannot be uplifted to D=5, namely:

\begin{itemize}
\item  $\mathcal{N}=2$ supergravity \textit{minimally coupled} to $n$ vector
multiplets \cite{Luciani:1977hp} (in 
\autoref{N=2-d=4-quadratic}). It has a quadratic U-invariant
polynomial, and it does \textit{not} enjoy a Jordan algebraic
formulation.

\item  $\mathcal{N}=3$ matter coupled supergravity \cite{Castellani:1985ka}
(in \autoref{N=3,d=4}). It has a quadratic U-invariant
polynomial, and it does \textit{not} enjoy a Jordan algebraic
formulation.

\item  $\mathcal{N}=5$ \textit{``pure''} supergravity \cite{deWit:1981yv}
(in  \autoref{N=5,d=4}). It enjoys a formulation in terms of $%
M_{2,1}\left( \mathds{O}\right) $, the Jordan triple system
generated by the $2\times 1$ vector over the octonions $\mathds{O}$
\cite {Gunaydin:1983rk,Gunaydin:1983bi}. Among the symmetric
supergravities with \textit{quartic} U-invariant polynomial, it
stands on a special footing, because its U-invariant polynomial is
a \textit{perfect square} when written in terms of the
scalar-dependent skew-eigenvalues of the $5\times 5$ complex
antisymmetric central charge matrix $Z_{AB}$. This property,
discussed in \cite{Ferrara:2008ap}, drastically simplifies the case
study of charge orbits.
\end{itemize}

For the convenience of the reader we summarize here our main original results together with where they appear in the text:

(1) In \autoref{sec:magic} the small (rank 3,2,1) orbits and moduli spaces of the magic $D=4, \mathcal{N}=2$ models based on degree-3 quaternionic, complex, real Jordan algebras are derived. The results are presented in the three $\alg=\R, \C, \Q$ sub-blocks of \autoref{tab:Mcharges}.  The $\alg=\Oct$ orbits as well as the large  $\alg=\R, \C, \Q$ orbits appearing in \autoref{tab:Mcharges} were previously obtained in \cite{Ferrara:1997uz}.
 In \autoref{sec:N6}  the $\mathcal{N}=2, D=4$ magic quaternionic case is compared to its ``twin'' $\mathcal{N}=6$ theory \cite{Andrianopoli:1997pn,Bellucci:2006xz,Ferrara:2008ap} and the supersymmetry analysis of twin  black hole charge orbits is carried out and presented in \eqref{eq:N6}.

(2) In \autoref{sec:N4N2} the small (rank 3,2,1) orbits and moduli spaces of the infinite sequences of  $D=4, \mathcal{N}=4$ and $D=4, \mathcal{N}=2$ Maxwell-Einstein theories are derived. The results are presented in \autoref{tab:n4redorbits} and \autoref{tab:n2redorbits}, respectively.  The  the large   orbits appearing in \autoref{tab:n4redorbits} and \autoref{tab:n2redorbits} were previously obtained in \cite{Ferrara:1997uz, Bellucci:2006xz, Andrianopoli:2006ub, Cerchiai:2009pi}.
In \autoref{sec:STU} it is observed that for the triality-symmetric $\mathcal{N}=2$  $STU$ model each of the rank 3 and rank 2 orbits split into two
\emph{isomorphic yet physically distinct} (BPS vs. non-BPS) orbits.

(4) In \autoref{sec:ST2} and \autoref{sec:T3} the small orbits and moduli spaces of the $ST^2$ and $T^3$ models are derived. For the $ST^2$ model the small orbits may be obtained from \autoref{tab:n2redorbits}  by setting $n=1$ (when this is still well defined - when it is not, the orbit is not present). The $T^3$ orbits are presented in  \autoref{tab:orbitsT3}.  It is established that while the BPS large orbit of the $T^3$ model (which one could think as
the simplest example of BPS-supporting charge orbit in $D=4, \mathcal{N}=2$ Maxwell-Einstein supergravity) has no continuous stabilizer it does in fact have a  $\mathds{Z}_3$ stabilizer.

(5) In \autoref{sec:N2mc}, \autoref{sec:N3} and \autoref{sec:N5} the unique small orbits and moduli spaces of the $\mathcal{N}=2$ minimally coupled, $\mathcal{N}=3$ matter coupled and
$\mathcal{N}=5$ pure supergravities are obtained, respectively.

\section{BH Charge Orbits in $D = 5$ Symmetric Supergravities}\label{sec:D5}

\subsection{Cubic Jordan Algebras \label{sec:J}}

A Jordan algebra $\mathfrak{J}$ is a vector space defined over a
ground field $\mathds{F}$ equipped with a bilinear product
satisfying
\begin{equation}\label{eq:Jid}
X\circ Y =Y\circ X,\quad X^2\circ (X\circ Y)=X\circ (X^2\circ Y), \quad\forall\ X, Y \in \mathfrak{J}.
\end{equation}
The class of \emph{cubic} Jordan algebras is constructed as follows
\cite{McCrimmon:2004}. Let $V$ be a vector space equipped with a
cubic norm, \textit{i.e.} an homogeneous map of degree three,
\[
N:V\to \mathds{F}, \quad\text{where}\quad N(\lambda
X)=\lambda^3N(X), \forall \lambda \in \mathds{F}, X\in V,
\]
such that
\begin{equation}
N(X, Y, Z):=\frac{1}{6}[N(X+ Y+ Z)-N(X+Y)-N(X+ Z)-N(Y+
Z)+N(X)+N(Y)+N(Z)]
\end{equation}
is trilinear.
If $V$ further contains a base point $N(c)=1, c\in V$  one may define the following three maps,
    \begin{equation}\label{eq:cubicdefs}
    \begin{split}
    \Tr:V\to\mathds{F};\quad& X    \mapsto3N(c, c, X),\\
         S: V\times V\to\mathds{F};\quad& (X,Y)       \mapsto6N(X, Y, c),\\
    \Tr:V\times V\to\mathds{F};\quad& (X,Y)      \mapsto\Tr(X)\Tr(Y)-S(X, Y).
    \end{split}
    \end{equation}

A cubic Jordan algebra $\mathfrak{J}$, with multiplicative identity
$\mathds{1}=c$, may be derived from any such vector space if $N$ is
\emph{Jordan cubic}. That is: if (i) the trace bilinear form
\eqref{eq:cubicdefs} is non-degenerate, and if (ii) the quadratic
adjoint map \be\label{eq:Jcubic}
\sharp\colon\mathfrak{J}\to\mathfrak{J}, \ee
 uniquely defined by
\be
\Tr(X^\sharp, Y) = 3N(X, X, Y),\ee
 satisfies
$(X^{\sharp})^\sharp=N(X)X$, $\forall X\in \mathfrak{J}$. The Jordan
product can then be implemented as follows: \be X\circ Y =
\half\big(X\times Y+\Tr(X)Y+\Tr(Y)X-S(X, Y)\mathds{1}\big), \ee
where, $X\times Y$ is the linearisation of the quadratic adjoint:
$X\times Y:= (X+Y)^\sharp-X^\sharp-Y^\sharp$.

The \emph{degree} of a cubic Jordan algebra is defined as the number
of linearly independent \emph{irreducible idempotents}:
\[
E\circ E=E, \quad \Tr(E)=1, \quad E\in \J.
\]
Two important symmetry groups, $\Aut(\J)$ and $\Str_0(\J)$, are
given by the set of $\F$-linear transformations preserving the
Jordan product and cubic norm, respectively. In particular,
$\Str_0(\J)$ is the U-duality group $G_5$ of the corresponding $D=5$
supergravity, and the corresponding vector multiplets' scalar
manifold is given by
\begin{equation}
\frac{\text{Str}_{0}\left( \frak{J}\right) }{\text{Aut}\left( \frak{J}%
\right) },
\end{equation}
which is isomorphic to the BPS rank 3 orbit in the symmetries theories with
8 supersymmetries - related to Jordan algebras -  in which
$\text{Aut}\left( \frak{J}\right)$ is the maximal compact subgroup
(\textit{mcs}) of $\text{Str}_{0}\left( \frak{J}\right)$, as well.

The conventional concept of matrix rank may be generalised to a cubic Jordan algebra in a natural and $\Str_0(\J)$ invariant manner. The rank of an arbitrary element $X\in\J$ is uniquely defined by \cite{Jacobson:1968}:
\be\label{eq:Jrank}
\begin{split}
\textrm{Rank} X =1& \Leftrightarrow X^\sharp=0;\\
\textrm{Rank} X =2& \Leftrightarrow N(X)=0,\;X^\sharp\not=0;\\
\textrm{Rank} X =3& \Leftrightarrow N(X)\not=0.\\
\end{split}
\ee

\subsection{$\mathcal{N}=8$\label{sec:D3N8}}

The $27=3+3\text{dim}_{\mathds{R}}\mathds{O}^{s}$ electric BH
charges may be represented as elements
\begin{equation}
Q=\begin{pmatrix}q_1&Q_s&\overline{Q_c}\\\overline{Q_s}&q_2&Q_v\\Q_c&\overline{Q}_v&q_3\end{pmatrix}, \quad \text{where} \quad q_1,q_2, q_3 \in \R \quad \text{and} \quad Q_{v,s,c}\in\sO
\end{equation}
of the $27$-dimensional  Jordan algebra $\JOs$ of $3\times 3$ Hermitian matrices over the split-octonions  $\sO$.  The cubic norm is  defined as,
\begin{equation}\label{eq:cubicnormexp}
N(Q)=q_1q_2q_3-q_1 Q_v\overline{Q_v}-q_2 Q_c\overline{Q_c}-q_3 Q_s \overline{Q_s} +(Q_vQ_c)Q_s+\overline{Q_s}(\overline{Q_c}\overline{Q_v}).
\end{equation}
One finds that the quadratic adjoint \eqref{eq:Jcubic} is given by
\begin{equation}\label{eq:quadadjexp}
Q^\sharp=\begin{pmatrix}q_2q_3-|Q_v|^2&\overline{Q_c}\overline{Q_v}-q_3
Q_s&Q_sQ_v-q_2 \overline{Q_c}\\Q_vQ_c-q_3\overline{Q_s}&q_1q_3-|Q_c|^2&\overline{Q_s}\overline{Q_c}-q_1 Q_v\\
\overline{Q_v}\overline{Q_s}-q_2
Q_c&Q_cQ_s-q_1\overline{Q_v}&q_1q_2-|Q_s|^2\end{pmatrix},
\end{equation}
from which it is derived that $Q\circ P=\half(QP+PQ)$. The cubic
Jordan algebra $\JOs$ has  irreducible idempotents
 given by \be E_1=\pmthree{1}{0}{0}{0}{0}{0}{0}{0}{0};\qquad
E_2=\pmthree{0}{0}{0}{0}{1}{0}{0}{0}{0};\qquad
E_3=\pmthree{0}{0}{0}{0}{0}{0}{0}{0}{1}. \ee The  $D=5$,
$\mathcal{N}=8$ U-duality group is given by  the reduced structure
group $\Str_0(\JOs)=E_{6(6)}$, which is the maximally non-compact
(split) form of $E_{6}(\C)$ under which $Q\in \JOs$ transforms as
the fundamental $\rep{27}$. The BH entropy is then given by (recall
Eq. (\ref{entropy-D=5}) \be S_{D=5,
\text{BH}}=\pi\sqrt{|I_3(Q)|}=\pi\sqrt{|N(Q)|}.
\label{entropy-D=5-II} \ee The U-duality charge orbits are
classified according to the $E_{6(6)}$-invariant Jordan rank of the
charge vector, as defined in \eqref{eq:Jrank}. This precisely
reproduces the classification originally obtained in
\cite{Ferrara:1997uz,Lu:1997bg}. The maximally split form of the
U-duality group, which corresponds to the use of the
split-octonions\footnote{The split-octonions are not division, but
are composition: $|ab|=|a||b|$.}, is the most powerful in the sense
that for each rank there is a \textit{unique} canonical form to
which all elements may be transformed. More precisely, we have the
following
\begin{theorem}\emph{\cite{Ferrara:1997uz,Krutelevich:2002}} Every BH charge vector $Q\in\JOs$ of a given rank is $E_{6(6)}$ related to one of the following canonical forms:
\begin{enumerate}
\item Rank 1
\begin{enumerate}
\item  $Q_{1}=(1,0,0)=E_1$
\end{enumerate}
\item Rank 2
\begin{enumerate}
\item   $Q_{2}=(1,1,0)=E_1+E_2$
\end{enumerate}
\item Rank 3
\begin{enumerate}
\item   $Q_{3}=(1,1,k)=E_1+E_2+kE_3$
\end{enumerate}
\end{enumerate}
\end{theorem}
The orbit stabilizers are summarized in \autoref{tab:d5N8orbits}. We
will see that the orbit structure of theories with less
supersymmetry is a progressive splitting of this exceptionally
simple case.
\begin{table}[!ht]
\small \caption{Charge orbits, corresponding \textit{moduli spaces}
and the number $\#$ of "non-flat" scalar directions of  $D=5,
\mathcal{N}=8$ supergravity defined over
$\JOs$ \cite{Ferrara:1997uz}.}\label{tab:d5N8orbits}
\begin{ruledtabular}
\begin{ruledtabular}\begin{tabular*}{\textwidth}{@{\extracolsep{\fill}}M{l}lM{l}*{3}{M{c}}}
\multicolumn{6}{c}{$\JOs$,  $M=E_{6(6)}/\Usp(8)$}                    \\[8pt]
\hline 
 \textrm{Rank}  &BH& \text{Susy}            & \text{Charge orbit}~  \mathcal{O}     & \text{Moduli space} ~ \mathcal{M} &      \#  \\[8pt]
\hline  
 1  &{small critical}&    1/2     &  \frac{E_{6(6)}}{\SO(5,5)\ltimes \R^{16}} &\frac{\SO(5,5)}{\SO(5)\times\SO(5)}\ltimes \R^{16}     & 1   \\[8pt]
 2&{small light-like}&    1/4     &  \frac{E_{6(6)}}{\SO(5,4)\ltimes \R^{16}} & \frac{\SO(5,4)}{\SO(5)\times\SO(4)}\ltimes \R^{16}    &   6 \\[8pt]
3&{large}&    1/8 &  \frac{E_{6(6)}}{F_{4(4)}}    &  \frac{F_{4(4)}}{\Usp(6)\times\SU(2)}   &         14            \\
\end{tabular*}\end{ruledtabular}
\end{ruledtabular}
\end{table}

\subsection{$\mathcal{N}=2$ Magic}
The $3+3\dim \alg$ electric BH charges may be represented as
elements
\begin{equation}
Q=\begin{pmatrix}q_1&Q_s&\overline{Q_c}\\\overline{Q_s}&q_2&Q_v\\Q_c&\overline{Q}_v&q_3\end{pmatrix}, \quad \text{where} \quad q_1,q_2, q_3 \in \R \quad \text{and} \quad Q_{v,s,c}\in\alg
\end{equation}
of the $(3+3\dim \alg)$-dimensional  Jordan algebra $\JA$ of
$3\times 3$ Hermitian matrices over the division algebra $\alg$
\cite{Gunaydin:1983bi}.  The irreducible idempotents, quadratic
adjoint and cubic norm are as in \autoref{sec:D3N8}. The magic
$D=5$, $\mathcal{N}=2$ U-duality groups $G_{5}^{\alg}$ are given by
the reduced structure group $\Str_0(\JA)$. For $\alg=\R, \C, \Q,
\Oct$ the U-duality $G_{5}^{\alg}$  is $\SL(3, \R), \SL(3, \C),
\SU^\star(6),E_{6(-26)}$ under which $Q\in \JA$ transforms as a
$\rep{6, 9, 15, 27}$, respectively. The BH entropy is given by Eq.
(\ref{entropy-D=5-II}). Once again, the U-duality charge orbits are
classified according to the $G_{5}^{\alg}$-invariant Jordan rank of
the charge vector. More precisely, we have the following
\begin{theorem} \emph{\cite{Ferrara:1997uz,Shukuzawa:2006}}
 Every BH charge vector $Q\in\JA$ of a given rank is $G_{5}^{\alg}$ related to one of the following canonical forms:
\begin{enumerate}
\item Rank 1
\begin{enumerate}
\item  $Q_{1a}=(1,0,0)=E_1$
\item $ Q_{1b}=(-1,0,0)=-E_1$
\end{enumerate}
\item Rank 2
\begin{enumerate}
\item   $Q_{2a}=(1,1,0)=E_1+E_2$
\item  $ Q_{2b}=(-1,1,0)=-E_1+E_2$
\item  $ Q_{2c}=(-1,-1,0)=-E_1-E_2$
\end{enumerate}
\item Rank 3
\begin{enumerate}
\item   $Q_{3a}=(1,1,k)=E_1+E_2+kE_3$
\item   $Q_{3b}=(-1,-1,k)=-E_1-E_2+kE_3$
\end{enumerate}
\end{enumerate}
\end{theorem}

Note, the orbits generated by the conical forms $Q_{1a}$ and
$Q_{1b}$ are isomorphic, as are those generated by $Q_{2a}$ and
$Q_{2c}$. The light-like 1/4-BPS orbit of the $\mathcal{N}=8$ splits
into one 1/2-BPS and one non-BPS orbit, as does the large 1/8-BPS
orbit. Note, the critical 1/2-BPS orbit remains intact
\cite{Cerchiai:2010xv}. The orbits are summarized in
\autoref{tab:d5magicorbits} (the exceptional - octonionic - case was
firstly derived in \cite{Ferrara:1997uz}). Note that the
$\mathcal{N}=2$ $\JH$ theory has a ``dual'' interpretation as
$\mathcal{N}=6$ supergravity,  as described in
\cite{Cerchiai:2010xv}.
\begin{table}[!ht]
\small \caption{Charge orbits, corresponding \textit{moduli spaces},
and number $\#$ of "non-flat" scalar directions of the magic $D=5,
\mathcal{N}=2$ supergravities defined over $\JA$, $\alg=\R, \C, \Q,
\Oct$ \cite{Ferrara:2006xx, Cerchiai:2010xv}.}\label{tab:d5magicorbits}
\begin{ruledtabular}\begin{tabular*}{\textwidth}{@{\extracolsep{\fill}}M{l}lM{l}*{3}{M{c}}}
\toprule
\toprule
\multicolumn{6}{c}{$\JO$, $n_V=26$, $M=E_{6(-26)}/F_{4(-52)}$}                                                                                                                                                                                                                                                                     \\[8pt]
 \textrm{Rank}  &BH& \text{Susy}            & \text{Charge orbit}~  \mathcal{O}     & \text{Moduli space} ~ \mathcal{M} &      \#   \\
\toprule
 1              &{small critical}&    1/2     &  \frac{E_{6(-26)}}{\SO(1,9)\ltimes \R^{16}}   &\frac{\SO(1,9)}{\SO(9)}\ltimes \R^{16}     & 1   \\[8pt]
 2a             &{small light-like}&  0       &  \frac{E_{6(-26)}}{\SO(1,8)\ltimes \R^{16}}   & \frac{\SO(1,8)}{\SO(8)}\ltimes \R^{16}    &   2 \\[8pt]
 2b             &{small light-like}&  1/2     &  \frac{E_{6(-26)}}{\SO(9)\ltimes \R^{16}}     & \R^{16}                                   &       10         \\[8pt]
3a  (k>0)       &{large}& 1/2                 &  \frac{E_{6(-26)}}{F_{4(-52)}}=M                & -    &        26            \\[8pt]
 3b  (k>0)      &{large}& 0\; {\scriptstyle (Z_H\not=0)}          &   \frac{E_{6(-26)}}{F_{4(-20)}}               & \frac{F_{4(-20)}}{\SO(9)} &   10                  \\[8pt]
 \toprule
\toprule
\multicolumn{6}{c}{$\JH$, $n_V=14$, $M=\SU^\star(6)/\Usp(6)$}                                                                                                                                                                                                                                                                     \\[8pt]
 \textrm{Rank}  &BH& \text{Susy}            & \text{Charge orbit}~  \mathcal{O}     & \text{Moduli space} ~ \mathcal{M}    &      \#            \\
\toprule
 1              &{small critical}&    1/2     &  \frac{\SU^\star(6)}{[\SO(1,5)\times \SO(3)]\ltimes \R^{(4, 2)}}  &\frac{\SO(1,5)}{\SO(5)}\ltimes \R^{(4, 2)} &    1      \\[8pt]
 2a             &{small light-like}&  0       &  \frac{\SU^\star(6)}{[\SO(1,4)\times \SO(3)]\ltimes \R^{(4, 2)}}  & \frac{\SO(1,4)}{\SO(4)}\ltimes \R^{(4, 2)}   &        2             \\[8pt]
 2b             &{small light-like}&  1/2     &  \frac{\SU^\star(6)}{[\SO(5)\times \SO(3)]\ltimes \R^{(4, 2)}}    & \R^{(4, 2)} &     6              \\[8pt]
3a  (k>0)       &{large}& 1/2 &  \frac{\SU^\star(6)}{\Usp(6)}=M             & -    &            14        \\[8pt]
 3b  (k>0)      &{large}& 0\; {\scriptstyle (Z_H\not=0)}          &   \frac{\SU^\star(6)}{\Usp(2,4)}              & \frac{\Usp(2,4)}{\Usp(2)\times \Usp(4)}   &           6            \\[8pt]
 \toprule
\toprule
\multicolumn{6}{c}{$\JC$, $n_V=8$, $M=\SL(3, \C)/\SU(3)$}                                                                                                                                                                                                                                                                     \\[8pt]
 \textrm{Rank}  &BH& \text{Susy}            & \text{Charge orbit}~  \mathcal{O}     & \text{Moduli space} ~ \mathcal{M}    &      \#            \\
\toprule
 1              &{small critical}&    1/2     & \frac{\SL(3, \C)}{[\SO(1,3)\times\SO(2)]\ltimes \R^{(2,2)}}   &\frac{\SO(1,3)}{\SO(3)}\ltimes \R^{(2, 2)} &     1     \\[8pt]
 2a             &{small light-like}&  0       & \frac{\SL(3, \C)}{[\SO(1,2)\times\SO(2)]\ltimes \R^{(2,2)}}   & \frac{\SO(1,2)}{\SO(2)}\ltimes \R^{(2, 2)} & 2  \\[8pt]
 2b             &{small light-like}&  1/2     & \frac{\SL(3, \C)}{[\SO(3)\times\SO(2)]\ltimes \R^{(2,2)}}     & \R^{(2, 2)} &     4              \\[8pt]
3a  (k>0)       &{large}& 1/2                 & \frac{\SL(3, \C)}{\SU(3)}=M                                     & -    &            8         \\[8pt]
 3b  (k>0)&{large}&   0\; {\scriptstyle (Z_H\not=0)}          & \frac{\SL(3, \C)}{\SU(1,2)}                                   & \frac{\SU(1,2)}{\U(1)\times\SU(2)}    &       4               \\[8pt]
  \toprule
\toprule
\multicolumn{6}{c}{$\JR$, $n_V=5$, $M=\SL(3, \R)/\SO(3)$}                                                                                                                                                                                                                                                                     \\[8pt]
 \textrm{Rank}  &BH& \text{Susy}            & \text{Charge orbit}~  \mathcal{O}     & \text{Moduli space} ~ \mathcal{M}    &      \#            \\
\toprule
 1              &{small critical}&    1/2     & \frac{\SL(3, \R)}{\SO(1,2)\ltimes \R^{2}}         &\frac{\SO(1,2)}{\SO(2)}\ltimes \R^{2} &         1 \\[8pt]
 2a             &{small light-like}&  0       & \frac{\SL(3, \R)}{\SO(1,1)\ltimes \R^{2}}         & \SO(1,1)\ltimes \R^{2} &  2 \\[8pt]
 2b             &{small light-like}&  1/2     & \frac{\SL(3, \R)}{\SO(2)\ltimes \R^{2}}           & \R^{2} &      3          \\[8pt]
3a  (k>0)       &{large}& 1/2                 & \frac{\SL(3, \R)}{\SO(3)}=M                         & -    &                   5   \\[8pt]
 3b  (k>0)&{large}&   0\; {\scriptstyle (Z_H\not=0)}          & \frac{\SL(3, \R)}{\SO(1,2)}                       & \frac{\SO(1,2)}{\SO(2)}   &                  3     \\[8pt]
\bottomrule
\bottomrule
\end{tabular*}\end{ruledtabular}
\end{table}

\subsection{The $\mathcal{N}=4$ and $\mathcal{N}=2$ Reducible Jordan Symmetric Sequences}

\subsubsection{$\mathcal{N}=4$}

For $\mathcal{N}=4$ supergravity coupled to $n_V$ vector multiplets,
the $n+5$ electric BH charges may be represented as elements ($\mu
:=0,I$, where $I=1,...,n+3$) \be Q=(q;q_\mu),\quad \text{where}\quad
q\in\R,\;  q_\mu \in \R^{5, n-1}, \ee of the $(n+5)$-dimensional
reducible cubic Jordan algebra $\Jnf$ (note that the index $0$
pertains to one of the 5 graviphotons). Note, we have adopted the
$(5, n-1)$ convention to emphasize the relation to the corresponding
$D=4$ theory, whereas in \cite{Cerchiai:2010xv} the $(5, n_V)$
convention was used, \textit{i.e.} $n=n_V+1$. The cubic norm is
defined as \be\label{eq:cubicnormN4} N(Q)=qq_\mu q^\mu, \ee where
the index $\mu$ has been raised with the $(+^5, -^{n-1})$ signature
metric $\eta^{\mu\nu}$; the positive signature pertains to the 5
graviphotons of the theory, whereas the negative one pertains to the
$n-1$ Abelian matter (vector) supermultiplets coupled to the gravity
multiplet. The reduced structure group reads \be
G_5=\Str_0(\J_{5,n-1})=\SO(1,1)\times\SO(5,n-1). \ee For
$\lambda\in\R, \Lambda\in\SO(5, n-1)$, its action on the charge
vector reads \be \label{jazz-2} (q;q_\mu)\mapsto (e^{2\lambda}q;
e^{-\lambda}\Lambda_{\mu}{}^{\nu}q_\nu). \ee One finds that the
quadratic adjoint \eqref{eq:Jcubic} is  given by,
\begin{equation}\label{eq:quadadjred}
Q^\sharp=(q_\mu q^\mu; qq_0, -qq_I),
\end{equation}
from which it is derived that\footnote{Note, this construction  appears to be undemocratic in the sense that it picks out one of the graviphotons $q_0$ as special. This is due to the undemocratic choice of base point $c=(1;1,0)$ we have used. This choice was made for convenience, but one could have equally used a ``democratic'' base point, valid for any signature $\J_{p,q}$ with $p\geq 1$, $c=(p^{-1}; 1,1,\ldots,1,0,0,\ldots, 0)$, which for $p=5$ treats all five graviphotons on the same footing. Of course, this is just a matter of conventions and the results are unaffected.} \be \label{jazz-1} Q\circ P=(qp;
q_0p_0-q_Jp^J, q_0p_I+p_0q_I), \ee where the index $I$ has been
raised with the $(+^4, -^{n-1})$ signature metric $\eta^{nm}$.
Consequently, the automorphism group is given by \be
\Aut(\Jnf)=\SO(4, n-1). \ee Three irreducible idempotents  are given by
\be\label{eq:N4idem} E_1=(1;0);\quad
E_2=(0;\half,0,0,0,0,\half,0,\ldots);\quad
E_3=(0;\half,0,0,0,0,-\half,0,\ldots). \ee The U-duality charge
orbits are classified according to the  $\SO(1,1)\times\SO(5, n-1)$
invariant Jordan \textit{rank} of the charge vector. More precisely,
the following theorem \cite{Borsten:2011nq} holds.
\begin{theorem} Every BH charge vector $Q=(q; q_\mu)\in\Jnf$ of a given rank is $\SO(1,1)\times\SO(5,n-1)$ related one of the following canonical forms:
\begin{enumerate}
\item Rank 1
\begin{enumerate}
\item  $Q_{1a}=E_1$
\item $ Q_{1b}=-E_1$
\item $ Q_{1c}=E_2$
\end{enumerate}
\item Rank 2
\begin{enumerate}
\item   $Q_{2a}=E_2+E_3$
\item  $ Q_{2b}=E_2-E_3$
\item  $ Q_{2c}=E_1+E_2$
\item  $ Q_{2d}=-E_1-E_2$
\end{enumerate}
\item Rank 3
\begin{enumerate}
\item   $Q_{3a}=E_1+E_2+kE_3$
\item   $Q_{3b}=-E_1+E_2+kE_3$
\end{enumerate}
\end{enumerate}
\end{theorem}
Note, the orbits $1a$ and $1b$ are physically equivalent, and have
isomorphic cosets. The same applies to $2c$ and $2d$. The orbits are
summarized in \autoref{tab:d5N4reducbleorbits} \cite{Cerchiai:2010xv}.
\begin{table}[!ht]
\caption{Charge orbits, corresponding \textit{moduli spaces} and
number $\#$ of ``non-flat" scalar directions of the reducible $D=5,
\mathcal{N}=4$ supergravities defined over
$\J_{5,n-1}=\R\oplus\Gamma_{5, n-1}$ \cite{Cerchiai:2010xv}. The
scalar manifold reads
$M=[\SO(1,1)\times\SO(5,n-1]/[\SO(5)\times\SO(n-1)]$, with
$\dim_\R=5n-4$}\label{tab:d5N4reducbleorbits}
\begin{ruledtabular}\begin{tabular*}{\textwidth}{@{\extracolsep{\fill}}M{l}cM{l}M{c}M{c}M{c}}
\toprule
 \toprule
\text{ Rank}&      BH               &\text{Susy}    & \text{Charge orbit}~ \mathcal{O}      & \text{Moduli space}~ \mathcal{M}  & \#    \\
\toprule
 1a         &\multirow{2}{*}{small critical}    &1/2    & \frac{\SO(1,1)\times\SO(5,n-1)}{\SO(5,n-1)}                               & \frac{\SO(5,n-1)}{\SO(5)\times\SO(n-1)}       &   1       \\[8pt]
 1c         &                   &1/2    & \frac{\SO(1,1)\times\SO(5,n-1)}{\SO(1,1)\times\SO(4, n-2)\ltimes \R^{4, n-2}}     &   \frac{\SO(1,1)\times\SO(4, n-2)}{\SO(4)\times\SO(n-2)}\ltimes {\scriptstyle\R^{4, n-2}}&   2    \\[8pt]
\toprule
 2a         &\multirow{3}{*}{small light-like}  &1/2    & \frac{\SO(1,1)\times\SO(5,n-1)}{\SO(4, n-1)}                                  &   \frac{\SO(4, n-1)} {\SO(4)\times\SO(n-1)}   &   n   \\[8pt]
 2b         &                   &0  & \frac{\SO(1,1)\times\SO(5,n-1)}{\SO(5, n-2) }                                 &   \frac{\SO(5, n-2)}{\SO(5)\times\SO(n-2)}        &     6    \\[8pt]
2c          &                   &1/4    & \frac{\SO(1,1)\times\SO(5,n-1)}{\SO(4, n-2)\ltimes\R^{4, n-2}}        &  \frac{\SO(4, n-2)}{\SO(4)\times\SO(n-2)}\ltimes{\scriptstyle \R^{4, n-2}}        &   2     \\[8pt]
\toprule
3ab(k>0)    &\multirow{2}{*}{large}     &1/4    & \frac{\SO(1,1)\times\SO(5,n-1)}{\SO(4, n-1)}                                &\frac{\SO(4, n-1)}{\SO(4)\times\SO(n-1)} &       n       \\[8pt]
 3b(k<0)    &                   &0\; {\scriptstyle (\hat{Z}_{AB, H}=0)} & \frac{\SO(1,1)\times\SO(5,n-1)}{\SO(5, n-2)}              & \frac{\SO(5, n-2)}{\SO(5)\times\SO(n-2)}      &   6           \\[8pt]
\bottomrule
\bottomrule
\end{tabular*}\end{ruledtabular}
\end{table}

\subsubsection{$\mathcal{N}=2$}

For $\mathcal{N}=2$ theories coupled to $n_V$ vector multiplets,
whose scalar manifolds belong to the so-called Jordan symmetric
sequence of the real special geometry, the $n+1$ electric BH charges
may be represented as elements ($\mu :=0,I$, where $I=1,...,n-1$)
\be Q=(q;q_\mu),\quad \text{where}\quad q\in\R,\;  q_\mu \in \R^{1,
n-1}, \ee of the $(n+1)$-dimensional reducible cubic Jordan algebra
$\Jnt$.  Once again, let us note that we have adopted the $(1, n-1)$
convention, in order to to emphasize the relation to the
corresponding $D=4$ theory, whereas in \cite{Cerchiai:2010xv} the
$(1, n_V)$ convention was used, \textit{i.e.} $n=n_V+1$. The set-up
and analysis is essentially as for the $\mathcal{N}=4$ case. The
principle difference is that the 1/4-BPS orbits split into one
1/2-BPS and one non-BPS orbit. This is captured in the connectedness
of the charge orbits \cite{Cerchiai:2010xv}, as we will discuss
below. This may be seen as a consequence of the Lorentzian nature of
$\Jnt$, contrasted to the genuine pseudo-Euclidean nature of $\Jnf$.
As for $\mathcal{N}=4$, the cubic norm is defined by
(\ref{eq:cubicnormN4}), but now the index $\mu$ is raised with the
$(+, -^{n-1})$ signature metric $\eta^{\mu\nu}$. The reduced
structure group is therefore \be
G_5=\Str_0(\J_{1,n-1})=\SO(1,1)\times\SO(1,n-1). \ee For
$\lambda\in\R, \Lambda\in\SO(1, n-1)$, its action on the charge
vector is given by Eq. (\ref{jazz-2}). Then, one finds that the
quadratic adjoint \eqref{eq:Jcubic} is  given by
\begin{equation}\label{eq:quadadjred}
Q^\sharp=(q_\mu q^\mu; qq^\mu),
\end{equation}
from which Eq. (\ref{jazz-1}) can be derived. Consequently, the
automorphism group is given by \be
\Aut(\Jnt)=\SO(n-1)=\text{\textit{mcs}}\left( \text{Str}_{0}\left(
\frak{J}_{1,n-1}\right) \right) . \ee
Three irreducible idempotents are given by
\be E_1=(1;0);\quad E_2=(0;\half, \half,0,\ldots);\quad
E_3=(0;\half,-\half,0,\ldots). \ee The U-duality charge orbits are
classified according to the  $\SO(1,1)\times\SO(1, n-1)$ invariant
Jordan \textit{rank} of the charge vector. More precisely, the
following theorem \cite{Borsten:2011nq} holds.
\begin{theorem} Every BH charge vector $Q=(q; q_\mu)\in\Jnt$ of a given rank is $\SO(1,1)\times\SO(1,n-1)$ related to one of the following canonical forms:
\begin{enumerate}
\item Rank 1
\begin{enumerate}
\item  $Q_{1a}=E_1$
\item $ Q_{1b}=-E_1$
\item $ Q_{1c}=E_2$
\end{enumerate}
\item Rank 2
\begin{enumerate}
\item   $Q_{2a}=E_2+E_3$
\item  $ Q_{2b}=E_2-E_3$
\item  $ Q_{2c}=E_1+E_2$
\item  $ Q_{2d}=-E_1-E_2$
\end{enumerate}
\item Rank 3
\begin{enumerate}
\item   $Q_{3a}=E_1+E_2+kE_3$
\item   $Q_{3b}=-E_1+E_2+kE_3$
\end{enumerate}
\end{enumerate}
Note, if one restricts to the identity-connected component of
$\SO(1,n-1)$, each of the orbits $Q_{1c}$, $Q_{2c}$  and $Q_{2d}$
splits into two cases, $Q^{\pm}_{1c}$, $Q^{\pm}_{2c}$ and
$Q^{\pm}_{2d}$, corresponding to the future and past light cones.
Similarly, $Q_{2a}$ splits into two disconnected components,
$Q^{\pm}_{2a}$, corresponding to the future and past hyperboloids.
For $k>0$ the orbits $Q_{3a}$ and $Q_{3b}$ also split into
disconnected future and past hyperboloids, $Q^{\pm}_{3a}$ and
$Q^{\pm}_{3b}$.
\end{theorem}

The orbits are summarized in \autoref{tab:d5reducbleorbits}. As
described in \cite{Cerchiai:2010xv}, the orbits  $Q^{\pm}_{2c}$,
$Q^{\pm}_{2d}$, $Q^{\pm}_{3a}$ and $Q^{\pm}_{3b}$  are BPS or
non-BPS according as the sign $+/-$ of $q$ is correlated or
anti-correlated, respectively, with the future/past branch on which
the orbit is defined.

\begin{table}[!ht]
\caption{Charge orbits, corresponding \textit{moduli spaces}, and
number $\#$ of ``non-flat'' scalar directions of the reducible $D=5,
\mathcal{N}=2$ supergravities defined over
$\J_{1,n-1}=\R\oplus\Gamma_{1, n-1}$ \cite{Cerchiai:2010xv}. The
scalar manifold reads $M=[\SO(1,1)\times\SO(1,n-1]/\SO(n-1)$, with
$\dim_\R M=n$.}\label{tab:d5reducbleorbits}
\begin{ruledtabular}\begin{tabular*}{\textwidth}{@{\extracolsep{\fill}}M{l}cM{l}M{c}M{c}M{c}}
\toprule
\toprule
\text{ Rank}&      BH       &\text{Susy}    & \text{Charge orbit}~ \mathcal{O}                                                          & \text{Moduli space}~ \mathcal{M}  & \#        \\
\toprule
 1a                 &\multirow{2}{*}{small critical}            &1/2    & \frac{\SO(1,1)\times\SO(1,n-1)}{\SO(1,n-1)}                               & \frac{\SO(1,n-1)}{\SO(n-1)}       &   1       \\[8pt]
 1c                 &           &1/2    & \frac{\SO(1,1)\times\SO(1,n-1)}{\SO(1,1)\times\SO(n-2)\ltimes \R^{n-2}}   &   \SO(1,1)\times\R^{n-2}                  &   2       \\[8pt]
\toprule
 2a                 &\multirow{6}{*}{small light-like}      &1/2    & \frac{\SO(1,1)\times\SO(1,n-1)}{\SO(n-1)}                                 &   -                                                           &   n   \\[8pt]
 2b                 &       &0      & \frac{\SO(1,1)\times\SO(1,n-1)}{\SO(1, n-2)}                              &   \frac{\SO(1, n-2)}{\SO(n-2)}        &     2    \\[8pt]
2c^{+}          &       &1/2    & \frac{\SO(1,1)\times\SO(1,n-1)}{\SO(n-2)\ltimes\R^{n-2}}      & \R^{n-2}                                              &   2     \\[8pt]
 2c ^{-}    &       &0      & \frac{\SO(1,1)\times\SO(1,n-1)}{\SO(n-2)\ltimes\R^{n-2}}      & \R^{n-2}                                              &   2          \\[8pt]
2d^{-}          &       &1/2    & \frac{\SO(1,1)\times\SO(1,n-1)}{\SO(n-2)\ltimes\R^{n-2}}      & \R^{n-2}                                              &   2          \\[8pt]
 2d ^{+}    &       &0      & \frac{\SO(1,1)\times\SO(1,n-1)}{\SO(n-2)\ltimes\R^{n-2}}      & \R^{n-2}                                              &   2          \\[8pt]
\toprule
3a^{+}(k>0)&\multirow{5}{*}{large}                          &1/2    & \frac{\SO(1,1)\times\SO(1,n-1)}{\SO(n-1)}                               & -                                                               &       n       \\[8pt]
3a^{-}(k>0) & &0\;  {\scriptstyle (Z_H\neq0)}  & \frac{\SO(1,1)\times\SO(1,n-1)}{\SO(n-1)}                               & -                                                               &       n           \\[8pt]
3b^{-}(k>0)&                            &1/2    & \frac{\SO(1,1)\times\SO(1,n-1)}{\SO(n-1)}                                     & -                                                             &   n           \\[8pt]
3b^{+}(k>0)&&0\;     {\scriptstyle (Z_H\not=0)} & \frac{\SO(1,1)\times\SO(1,n-1)}{\SO(n-1)}                                     & -                                                             &       n               \\[8pt]
3ab (k<0)  &        &0\; {\scriptstyle (Z_H\not=0)}
        & \frac{\SO(1,1)\times\SO(1,n-1)}{\SO(1, n-2)}                              &   \frac{\SO(1, n-2)}{\SO(n-2)}        &   2         \\[8pt]
\bottomrule
\bottomrule
\end{tabular*}\end{ruledtabular}
\end{table}

The \textit{non-Jordan symmetric} sequence \cite{deWit:1992cr}
\begin{equation}
M_{nJ,5,n}\equiv \frac{SO\left( 1,n\right) }{SO\left( n\right)
},~n=n_{V}\in \mathds{N},  \label{non-Jordan-symm}
\end{equation}
($n_{V}$ being the number of Abelian vector supermultiplets coupled to the $%
\mathcal{N}=2$, $D=5$ supergravity one) is the only (sequence of)
symmetric \textit{real special geometry} which is \textit{not}
related to a cubic Jordan algebra.
It is usually denoted by
$L\left( -1,n-1\right) $ in the classification of homogeneous
Riemannian $d$-spaces (see \textit{e.g.} \cite{deWit:1992wf}, and
Refs. therein).

As discussed in \cite{deWit:1992cr}, the isometries of the symmetric
real special space (\ref{non-Jordan-symm}) are not all contained in
the invariance group of the corresponding supergravity theory,
despite the fact that the latter group still acts transitively on
the space. By using the parametrization introduced in the last Sec.
of \cite{deWit:1991nm} and comparing \textit{e.g.} Eq. (5.1) of
\cite{deWit:1992wf} to Eq. (7) of \cite {deWit:1992cr}, it is
immediate to conclude that the $D=5$, $\mathcal{N}=2$
Maxwell-Einstein supergravity theory whose scalar manifold is given
by (\ref {non-Jordan-symm}) can be uplifted to a $D=6$, $(1,0)$
supergravity theory
with $n-1$ vector multiplets, but \textit{no tensor multiplets at all} ($%
n_{T}=0$). Thus, in absence of matter fields charged under a
non-trivial gauge group, the gravitational anomaly-free condition
implies that \cite {Salam:1985mi,RandjbarDaemi:1985wc} $n_{H}=272+n$
hypermultiplets must be coupled to the theory. On the other hand,
this theory is known not to satisfy the condition of conservation of
the gauge vector current (required by the consistency of the gauge
invariance \cite
{Ferrara:1996wv,Andrianopoli:2004xu,Ferrara:1997gh,Riccioni:1998th,Nishino:1997ff}%
); therefore, it seemingly has a $D=6$ uplift to $(1,0)$ chiral
supergravity which is \textit{not} anomaly-free, unless it is
embedded in a model where a non-trivial gauge group is present, with
charged matter (see \textit{e.g.}
\cite{Angelantonj:2002ct,Antoniadis:1997nz}).

We will not further considered this theory in the present
investigation, because it does not correspond to symmetric spaces in
$D=4$ \cite{deWit:1992cr}.

\section{BH Charge Orbits in $D = 4$ Symmetric Supergravities}\label{sec:D4}

\subsection{The Freudenthal Triple System\label{sec:FTS}}

Given a cubic Jordan algebra $\mathfrak{J}$ defined over a field
$\mathds{F}$, one is able to construct a FTS by defining the vector
space $\mathfrak{F(J)}(:=\mathfrak{F})$,
\begin{equation}
\mathfrak{F(J)}=\mathds{F\oplus F}\oplus \mathfrak{J\oplus J}.
\end{equation}
An arbitrary element $x\in \mathfrak{F(J)}$ may be written as a
formal ``$2\times 2$ matrix'',
\begin{equation}
x=\begin{pmatrix}\alpha&X\\Y&\beta\end{pmatrix} \quad\text{where} ~\alpha, \beta\in\mathds{F}\quad\text{and}\quad X, Y\in \mathfrak{J}.
\end{equation}
The FTS comes equipped with a non-degenerate bilinear antisymmetric quadratic form, a quartic form and a trilinear triple product \cite{Freudenthal:1954,Brown:1969}:
\begin{subequations}
\begin{enumerate}
\item Quadratic form $ \{x, y\}$: $\mathfrak{F}\times \mathfrak{F}\to \mathds{F}$
    \begin{equation}\label{eq:bilinearform}
    \{x, y\}=\alpha\delta-\beta\gamma+\Tr(X,W)-\Tr(Y,Z),\quad
    \textrm{where}\quad x=\begin{pmatrix}\alpha&X\\Y&\beta\end{pmatrix},\; y=\begin{pmatrix}\gamma&Z\\W&\delta\end{pmatrix}.
      \end{equation}
\item Quartic form $q:\mathfrak{F}\to \mathds{F}$
    \begin{equation}\label{eq:quarticnorm}
        q(x)=-2[\alpha\beta-\Tr(X, Y)]^2
         -8[\alpha N(X)+\beta N(Y)-\Tr(X^\sharp, Y^\sharp)].
        \end{equation}
\item Triple product $T:\mathfrak{F\times F\times F\to F}$ which is uniquely defined by
    \begin{equation}\label{eq:tripleproduct}
    \{T(x, y, w), z\}=q(x, y, w, z)
    \end{equation}
    where $q(x, y, w, z)$ is the full linearisation of $q(x)$ such that $q(x, x, x, x)=q(x)$.
\end{enumerate}
\end{subequations}
The \emph{automorphism} group is given by the set of invertible
$\F$-linear transformations preserving the quadratic and quartic
forms \cite{Freudenthal:1954,Brown:1969}, \be
\AutF{\mathfrak{F}}:=\{\sigma\in \Iso_{\F}(\mathfrak{F})| q(\sigma
x)=q(x), \{\sigma x, \sigma y\}=\{x,y\}, \;\forall x,y\in
\mathfrak{F}\}=\text{Conf}\left( \frak{J}\right). \ee Generally, the
automorphism group corresponds to the U-duality group of
corresponding 4-dimensional supergravities (see for example
\cite{Bellucci:2006xz,Rios:2007qn,Borsten:2010aa,Rios:2010br}, and Refs. therein). The
conventional concept of matrix rank may be generalised to
Freudenthal triple systems in a natural and $\Aut(\mathfrak{F})$
invariant manner. The rank of an arbitrary element
$x\in\mathfrak{F}$ is uniquely defined by \cite{Ferrar:1972,
Krutelevich:2004}: \be\label{eq:FTSrank}
\begin{split}
\textrm{Rank} x=1&\Leftrightarrow 3T(x,x,y)+x\{x,y\}x=0\;\forall y;\\
\textrm{Rank} x=2&\Leftrightarrow \exists y\;\textrm{s.t.}\;3T(x,x,y)+x\{x,y\}x\not=0,\;T(x,x,x)=0;\\
\textrm{Rank} x=3&\Leftrightarrow T(x,x,x)\not=0,\;q(x)=0;\\
\textrm{Rank} x=4&\Leftrightarrow q(x)\not=0.\\
\end{split}
\ee
\subsection{$\mathcal{N}=8$}

The $(1+27)+(1+27)$ electric$+$magnetic BH charges may be
represented as elements
\begin{equation}
x=\begin{pmatrix}-q_0&P\\Q&p^0\end{pmatrix}, \quad\text{where} \quad
p^0, q^0\in\R\quad\text{and}\quad Q, P\in\JOs
\end{equation}
of the Freudenthal triple system
$\mathfrak{F}^\alg:=\mathfrak{F}(\JOs)$.  The details may be found
in \autoref{sec:FTS} of \cite{Borsten:2011nq}, and in Refs. therein.
The automorphism group $\Aut(\FOs)\cong\text{Conf}\left(
\frak{J}_{3}^{\mathds{O}^{s}}\right)=E_{7(7)}$ is the  $D=4$,
$\mathcal{N}=8$ U-duality group, where $x\in \FA$ transforms as the
fundamental $\rep{56}$. The BH entropy is given by Eq.
(\ref{BH-entropy-D=4}), where $I_4(x)=\Delta(x)=\half q(x)$ is Cartan's
unique quartic invariant polynomial of $E_{7(7)}$
\cite{Kallosh:1996uy}. The U-duality charge orbits are classified
according to the $E_{7(7)}$-invariant FTS \textit{rank} of the
charge vector, as defined in \eqref{eq:FTSrank}. This reproduces the
classification originally obtained in
\cite{Ferrara:1997uz,Lu:1997bg}. More precisely, we have the
following

\begin{theorem}\label{thm:N8} \emph{\cite{Ferrara:1997uz,Krutelevich:2004,Borsten:2008wd}} Every BH charge vector $x\in\FOs$ of a given rank is $E_{7(7)}$ related one of the following canonical forms:
\begin{enumerate}
\item Rank 1
\begin{enumerate}
\item  $x_{1}=\begin{pmatrix}1&0\\0&0\end{pmatrix}$
\end{enumerate}
\item Rank 2
\begin{enumerate}
\item   $x_{2}=\begin{pmatrix}1&(1,0,0)\\0&0\end{pmatrix}$
\end{enumerate}
\item Rank 3
\begin{enumerate}
\item   $x_{3}=\begin{pmatrix}1&(1,1,0)\\0&0\end{pmatrix}$
\end{enumerate}
\item Rank 4
\begin{enumerate}
\item   $x_{4a}=k\begin{pmatrix}1&(-1,-1,-1)\\0&0\end{pmatrix}$
\item   $x_{4b}=k\begin{pmatrix}1&(1,1,1)\\0&0\end{pmatrix}$
\end{enumerate}
where $k>0$.
\end{enumerate}
\end{theorem}

As  anticipated, there is one orbit per rank, but with rank 4 splitting into $4a$ $(\Delta>0)$ 1/8-BPS and $4b$ $(\Delta<0)$ non-BPS. The orbits are summarized in \autoref{tab:D4N8}.

\begin{table}[!ht]
\caption{Charge orbits, \textit{moduli spaces}, and number $\#$ of
``non-flat" scalar directions of  $D=4, \mathcal{N}=8$ supergravity
defined over $\FOs$. $M=E_{7(7)}/\SU(8)$,
$\dim_\R=70$ \cite{Ferrara:1997uz}.}\label{tab:D4N8}
\begin{ruledtabular}\begin{tabular*}{\textwidth}{@{\extracolsep{\fill}}M{l}cM{l}M{c}M{c}M{c}}
\toprule
 \toprule
\text{ Rank}&      BH               &\text{Susy}    & \text{Charge orbit}~ \mathcal{O}      & \text{Moduli space}~ \mathcal{M}  & \#    \\
\toprule
 1      &doubly critical    &1/2    & \frac{E_{7(7)}}{E_{6(6)}\ltimes \R^{27}}              & \frac{E_{6(6)}}{\Usp(8)} \ltimes \R^{27}      &   1       \\[8pt]
\toprule
 2      &critical   &1/4    & \frac{E_{7(7)}}{\SO(6, 5)\ltimes \R^{32}\times \R}            &   \frac{\SO(6, 5)} {\SO(6)\times\SO(5)}\ltimes \R^{32}\times \R   &   7   \\[8pt]
\toprule
 3      &light-like &1/8    & \frac{E_{7(7)}}{F_{4(4)}\ltimes \R^{26}}                  &   \frac{F_{4(4)}} {\Usp(6)\times\SU(2)}\ltimes \R^{26}    &   16      \\[8pt]
 \toprule
4 (\Delta>0)    &\multirow{2}{*}{large}     &1/8    & \frac{E_{7(7)}}{E_{6(2)}}                               &\frac{E_{6(2)}}{\SU(6)\times\SU(2)}  &       30      \\[8pt]
 4 (\Delta<0)   &                   &0  & \frac{E_{7(7)}}{E_{6(6)}}             & \frac{E_{6(6)}}{\Usp(8)}      &   28              \\[8pt]
\bottomrule
\bottomrule
\end{tabular*}\end{ruledtabular}
\end{table}
\subsection{$\mathcal{N}=2$ Magic}\label{sec:magic}

The $(4+3\dim \alg)+(4+3\dim \alg)$ electric$+$magnetic BH charges
may be represented as elements
\begin{equation}
x=\begin{pmatrix}-q_0&P\\Q&p^0\end{pmatrix}, \quad\text{where} \quad
p^0, q^0\in\R\quad\text{and}\quad Q, P\in\JA
\end{equation}
of the Freudenthal triple system
$\mathfrak{F}^\alg:=\mathfrak{F}(\JA)$. The details may be found in
\autoref{sec:FTS}, Ref. \cite{Borsten:2011nq}, and in Refs. therein. The
magic $D=4$, $\mathcal{N}=2$ U-duality groups $G_{4}^{\alg}$ are
given by the automorphism group $\Aut(\FA)\cong\text{Conf}\left(
\frak{J}_{3}^{\mathds{A}}\right)$. For $\alg=\R, \C, \Q, \Oct$ the
U-duality group $G_{4}^{\alg}$ is $\Sp(6, \R), \SU(3, 3),
\SO^\star(12),E_{7(-25)}$. The $(8+6\dim \alg)$ charges transform
linearly as the threefold antisymmetric traceless tensor
$\rep{14}'$, the threefold antisymmetric self-dual tensor
$\rep{20}$, the chiral spinor $\rep{32}$ and the fundamental
$\rep{56}$ of   $\Sp(6, \R)$, $\SU(3,3)$, $\SO^\star(12)$ and
$E_{7(-25)}$, respectively.

The BH entropy is given by Eq. (\ref{BH-entropy-D=4}), where
$I_4(x)=\Delta(x)=\half q(x)$ is the unique quartic invariant polynomial
of $G_{4}^{\alg}$. The U-duality charge orbits are classified
according to the $G_{4}^{\alg}$-invariant FTS \textit{rank} of the
charge vector, as defined in \eqref{eq:FTSrank}. More precisely, we
have the following

\begin{theorem}\label{thm:Shuk} \emph{\cite{Ferrara:1997uz,Shukuzawa:2006}} Every BH charge vector $x\in\FA$ of a given rank is $G_{4}^{\alg}$ related one of the following canonical forms:
\begin{enumerate}
\item Rank 1
\begin{enumerate}
\item  $x_{1}=\begin{pmatrix}1&0\\0&0\end{pmatrix}$
\end{enumerate}
\item Rank 2
\begin{enumerate}
\item   $x_{2a}=\begin{pmatrix}1&(1,0,0)\\0&0\end{pmatrix}$
\item  $ x_{2b}=\begin{pmatrix}1&(-1,0,0)\\0&0\end{pmatrix}$
\end{enumerate}
\item Rank 3
\begin{enumerate}
\item   $x_{3a}=\begin{pmatrix}1&(1,1,0)\\0&0\end{pmatrix}$
\item   $x_{3b}=\begin{pmatrix}1&(-1,-1,0)\\0&0\end{pmatrix}$
\end{enumerate}
\item Rank 4
\begin{enumerate}
\item   $x_{4a}=k\begin{pmatrix}1&(-1,-1,-1)\\0&0\end{pmatrix}$
\item       $x_{4b}=k\begin{pmatrix}1&(1,1,-1)\\0&0\end{pmatrix}$
\item   $x_{4c}=k\begin{pmatrix}1&(1,1,1)\\0&0\end{pmatrix}$
\end{enumerate}
where $k>0$.
\end{enumerate}
\end{theorem}
Here, we see that the rank 2 and 3 orbits of the $\mathcal{N}=8$
theory split in to one 1/2-BPS orbit and one non-BPS orbit each. The
splitting of the large BHs is a little more subtle
\cite{Bellucci:2006xz}. There is, as always for $\mathcal{N}=2$, one
1/2-BPS ($I_4>0$) orbit, which we label $4a$. However, there is also
one non-BPS orbit for $I_4>0$, which has vanishing central charge at
the horizon $Z_H=0$. Finally, there is the universal non-BPS
$I_4<0$, which has non-vanishing central charge at the horizon. The
orbit stabilizers are summarized in \autoref{tab:Mcharges}. The
exceptional octonionic case is given as a detailed example in
\autoref{sec:orbitsmagicFTS}, which thus provides an alternative
derivation of the result obtained in \cite{Ferrara:1997uz}.

\subsubsection{$\mathcal{N}=2$ Magic Quaternionic \textit{versus} $\mathcal{N%
}=6$}\label{sec:N6}

As is well known \cite{Andrianopoli:1997pn,Bellucci:2006xz,Ferrara:2008ap}, $%
\mathcal{N}=2$ magic quaternionic and $\mathcal{N}=6$ supergravity share the
very same bosonic sector; they are both related to the simple, rank-$3$
Jordan algebra $\mathfrak{J}_{3}^{\mathbb{H}}$ over the quaternions, and
their scalar manifold is the rank-$3$ symmetric coset $\frac{SO^{\ast }(12)}{%
U(6)}$.

It should also be noticed that the two real, non-compact forms \ of $E_{7}$
given by $E_{7(7)}$ and $E_{7(-25)}$ contain $SO^{\ast }(12)\times SU(2)$ as
a maximal subgroup, and indeed both manifolds $\frac{E_{7(-25)}}{E_{6}\times
U(1)}$ (rank-$3$ special K\"{a}hler, with dim$_{\mathbb{C}}=27$) and $\frac{%
E_{7(7)}}{SU(8)}$ (rank-$7$, with dim$_{\mathbb{R}}=70$) contain the coset
space $\frac{SO^{\ast }(12)}{U(6)}$ as a submanifold. Such an observation
reveals the \textit{dual} role of the manifold $\frac{SO^{\ast }(12)}{U(6)}$%
: it is at the same time the $\sigma $-model scalar manifold of $\mathcal{N}%
=6$ supergravity and of $\mathcal{N}=2$ magic quaternionic Maxwell-Einstein
supergravity.

Starting from $\mathcal{N}=8$, the supersymmetry truncation down to $%
\mathcal{N}=6$ goes as follows:
\begin{gather}
\mathcal{N}=8:\left[ \left( 2\right) ,~8\left( \frac{3}{2}\right) ,~28\left(
1\right) ,~56\left( \frac{1}{2}\right) ,~70\left( 0\right) \right] ~\text{%
gravity~mult.}  \notag \\
\downarrow  \notag \\
\mathcal{N}=6:\left\{
\begin{array}{l}
\left[ \left( 2\right) ,~6\left( \frac{3}{2}\right) ,~16\left( 1\right)
,~26\left( \frac{1}{2}\right) ,~30\left( 0\right) \right] ~\text{%
gravity~mult.} \\
~ \\
2~\left[ \left( \frac{3}{2}\right) ,~6\left( 1\right) ,~15\left( \frac{1}{2}%
\right) ,~20\left( 0\right) ,\right] ~\text{gravitino~mults.}%
\end{array}%
\right.  \label{N=6}
\end{gather}%
In order to truncate the two $\mathcal{N}=6$ gravitino multiplets away, one
has to consider the $U$-duality branching for vectors reads
\begin{eqnarray}
E_{7\left( 7\right) } &\supset &SO^{\ast }\left( 12\right) \times SU\left(
2\right) ;  \notag \\
\mathbf{56} &=&\left( \mathbf{32},\mathbf{1}\right) +\left( \mathbf{12},%
\mathbf{2}\right) ,
\end{eqnarray}%
implying the truncation condition
\begin{equation}
SO^{\ast }\left( 12\right) \times SU\left( 2\right) :\left( \mathbf{12},%
\mathbf{2}\right) =0,  \label{trunc-1}
\end{equation}%
as well as the $\mathcal{R}$-symmetry branching (omitting $U(1)$ charges)
\begin{eqnarray}
\overset{\mathcal{N}=8~\mathcal{R}\text{-symmetry}}{SU\left( 8\right) }
&\supset &\overset{\mathcal{N}=6~\mathcal{R}\text{-symmetry}}{U\left(
6\right) }\times ~SU\left( 2\right) ;  \label{R-br} \\
\mathbf{8} &=&\left( \mathbf{6,1}\right) +\left( \mathbf{1},\mathbf{2}%
\right) ;  \notag \\
\mathbf{28} &=&\left( \mathbf{15},\mathbf{1}\right) +\left( \mathbf{1},%
\mathbf{1}\right) +\left( \mathbf{6},\mathbf{2}\right) ;  \notag \\
\mathbf{56} &=&\left( \mathbf{20},\mathbf{1}\right) +\left( \mathbf{6},%
\mathbf{1}\right) +\left( \mathbf{15},\mathbf{2}\right) ;  \notag \\
\mathbf{70} &=&\left( \mathbf{15},\mathbf{1}\right) +\left( \mathbf{15},%
\mathbf{1}\right) +\left( \mathbf{20},\mathbf{2}\right) ,  \notag
\end{eqnarray}%
implying the truncation conditions
\begin{equation}
U\left( 6\right) \times ~SU\left( 2\right) :\left( \mathbf{1},\mathbf{2}%
\right) =\left( \mathbf{6},\mathbf{2}\right) =\left( \mathbf{15},\mathbf{2}%
\right) =\left( \mathbf{20},\mathbf{2}\right) =0.  \label{trunc-2}
\end{equation}%
Note that the commuting $SU\left( 2\right) $ factor in (\ref{R-br}) may be
regarded as the \textquotedblleft extra\textquotedblright\ $\mathcal{R}$%
-symmetry truncated away in the supersymmetry reduction $\mathcal{N}%
=8\rightarrow \mathcal{N}=6$ obtained by imposing (\ref{trunc-1}) and (\ref%
{trunc-2}), which corresponds to the following scalar manifold embedding:
\begin{equation}
\frac{E_{7\left( 7\right) }}{SU\left( 8\right) }\supset \frac{SO^{\ast
}\left( 12\right) }{U\left( 6\right) }.  \label{s-emb}
\end{equation}

On the other hand, the supersymmetry truncation $\mathcal{N}=8\rightarrow
\mathcal{N}=2$ goes as follows:
\begin{gather}
\mathcal{N}=8:\left[ \left( 2\right) ,~8\left( \frac{3}{2}\right) ,~28\left(
1\right) ,~56\left( \frac{1}{2}\right) ,~70\left( 0\right) \right] ~\text{%
gravity~mult.}  \notag \\
\downarrow   \notag \\
\mathcal{N}=2:\left\{
\begin{array}{l}
\left[ \left( 2\right) ,~2\left( \frac{3}{2}\right) ,~\left( 1\right) \right]
~\text{gravity~mult.} \\
~ \\
6~\left[ \left( \frac{3}{2}\right) ,~2\left( 1\right) ,~\left( \frac{1}{2}%
\right) \right] ~\text{gravitino~mults.} \\
~ \\
15~\left[ \left( 1\right) ,~2\left( \frac{1}{2}\right) ,~2\left( 0\right) %
\right] ~\text{vector~mults.} \\
~ \\
10~\left[ 2\left( \frac{1}{2}\right) ,~4\left( 0\right) \right] ~\text{%
hypermults.}%
\end{array}%
\right.
\end{gather}%
In order to truncate the six $\mathcal{N}=2$ gravitino multiplets away, the
same condition (\ref{trunc-1}) on $U$-irreps. has to be imposed. On the
other hand, by reconsidering (\ref{R-br}) with the the different
interpretation of $\mathcal{R}$-symmetry branching $\mathcal{N}=8\rightarrow
\mathcal{N}=2$ (the commuting $SU\left( 6\right) $ factor in (\ref{R-br})
now refers to the \textquotedblleft extra\textquotedblright\ $\mathcal{R}$%
-symmetry truncated away), the following truncation conditions, different
from (\ref{trunc-2}), are obtained:
\begin{equation}
U\left( 6\right) \times ~SU\left( 2\right) :\left( \mathbf{6},\mathbf{1}%
\right) =\left( \mathbf{6},\mathbf{2}\right) =0.  \label{trunc-3}
\end{equation}%
Thus, by imposing (\ref{trunc-1}) and (\ref{trunc-3}), one achieves a
consistent truncation of $\mathcal{N}=8$ down to $\mathcal{N}=2$ magic
octonionic supergravity coupled to $15$ vector multiplets and $10$
hypermultiplets, which at the level of the scalar manifold reads:
\begin{equation}
\frac{E_{7\left( 7\right) }}{SU\left( 8\right) }\supset \frac{SO^{\ast
}\left( 12\right) }{U\left( 6\right) }\times \frac{E_{6\left( 2\right) }}{%
SU\left( 6\right) \times SU\left( 2\right) }.
\end{equation}%
The $\mathcal{N}=2$ hyper sector can be consistently truncated away, by
further imposing
\begin{equation}
U\left( 6\right) \times ~SU\left( 2\right) :\left( \mathbf{20},\mathbf{1}%
\right) =\left( \mathbf{20},\mathbf{2}\right) =0,
\end{equation}%
thus yielding (\ref{s-emb}).

On the other hand, starting from the $\mathcal{N}=2$ exceptional magic
supergravity with no hypermultiplets, the truncation down to its $\mathcal{N}%
=2$ magic quaternionic sub-theory is dictated by the following branchings ($H
$ is the local symmetry group of the scalar manifold, up to a $U\left(
1\right) $ factor):
\begin{gather}
U\text{-duality}:\left\{
\begin{array}{l}
E_{7\left( -25\right) }\supset SO^{\ast }\left( 12\right) \times SU\left(
2\right) , \\
\mathbf{56}=\left( \mathbf{32},\mathbf{1}\right) +\left( \mathbf{12},\mathbf{%
2}\right) ;%
\end{array}%
\right.  \\
H\text{-symmetry}:\left\{
\begin{array}{l}
E_{6\left( -78\right) }\supset SU\left( 6\right) \times SU\left( 2\right) ,
\\
\mathbf{27}=\left( \overline{\mathbf{6}},\mathbf{2}\right) +\left( \mathbf{15%
},\mathbf{1}\right) ,%
\end{array}%
\right.
\end{gather}%
implying the truncation conditions
\begin{eqnarray}
SO^{\ast }\left( 12\right) \times SU\left( 2\right)  &:&\left( \mathbf{12},%
\mathbf{2}\right) =0; \\
SU\left( 6\right) \times SU\left( 2\right)  &:&\left( \overline{\mathbf{6}},%
\mathbf{2}\right) =0.
\end{eqnarray}%
Under such positions, one achieves a consistent truncation of $\mathcal{N}=2$
exceptional Maxwell-Einstein supergravity down to its $\mathcal{N}=2$ magic
quaternionic sub-theory which at the level of the scalar manifold reads:
\begin{equation}
\frac{E_{7\left( -25\right) }}{E_{6\left( -78\right) }\times U\left(
1\right) }\supset \frac{SO^{\ast }\left( 12\right) }{U\left( 6\right) }.
\end{equation}

Once their origin as truncation has been clarified, it is thus evident that $%
\mathcal{N}=2$ quaternionic and $\mathcal{N}=6$, $D=4$ supergravities
exhibit \textit{indistinguishable} bosonic sectors, and therefore their
charge orbits are the same, and their attractor equations \cite%
{Bellucci:2006xz} have the same solutions.

In order to elucidate the different supersymmetry properties of the charge
orbits, by recalling the spin content of the $\mathcal{N}=6$ gravity
multiplet, it should be noticed that its $16$ vector fields decompose as $%
\mathbf{15}+\mathbf{1}$ with respect to the $\mathcal{N}=6$ $\mathcal{R}$%
-symmetry (as well as the $26$ gauginos and the $30$ scalar fields decompose
as $\mathbf{20}+\mathbf{6}$ and $\mathbf{15}+\overline{\mathbf{15}}$,
respectively). Thus, the $\mathcal{N}=6$ dyonic charge vector $\mathcal{Q}$
splits as
\begin{equation}
\mathcal{N}=6:\mathcal{Q}=\left( X,~Z_{AB},~\overline{Z}^{AB},~\overline{X}%
\right) ,  \label{C-N=6}
\end{equation}%
where $X$ is a \textit{complex} $SU(6)$-singlet, and $Z_{AB}$ ($A=1,...,6$)
is the complex $6\times 6$ antisymmetric central charge matrix. The
intertwining supersymmetry-preserving properties for the \textit{%
\textquotedblleft twin"} theories $\mathcal{N}=2$ magic quaternionic \textit{%
versus} \textquotedblleft pure" $\mathcal{N}=6$ can be obtained by noticing
that the $\mathcal{N}=2$ counterpart of (\ref{C-N=6}) is given by%
\begin{equation}
\mathcal{N}=2:\mathcal{Q}=\left( Z,~Z_{i},~\overline{Z}_{\overline{i}},~%
\overline{Z}\right) ,  \label{C-N=2}
\end{equation}%
where $Z_{i}\equiv D_{i}Z$ are the so-called \textit{matter charges}
(namely, the K\"{a}hler-covariant derivatives of the $\mathcal{N}=2$ central
charge $Z$). As summarized in Table 9 of \cite{Bellucci:2006xz}, (\ref{C-N=6}%
) and (\ref{C-N=2}) imply that the role of \textquotedblleft large" BPS
orbits and non-BPS orbits with (all) central charge(s) vanishing is \textit{%
flipped} under the \textit{exchange} $\mathcal{N}=2\longleftrightarrow
\mathcal{N}=6$; as mentioned, such a kind of \textit{\textquotedblleft
cross-symmetry\textquotedblright } is easily understood when noticing that
the $\mathcal{N}=2$ central charge $Z$ corresponds to the $SU(6)$-singlet
component $X$ of $\mathcal{Q}$ (\ref{C-N=6}), and that the $15$ complex $%
\mathcal{N}=2$ matter charges $Z_{i}$ correspond to the $15$ independent
complex elements of the $6\times 6$ antisymmetric $\mathcal{N}=6$ central
charge matrix $Z_{AB}$.

These considerations can be extended to \textquotedblleft small" charge
orbits, by observing that orbits with representatives having $Z=0$
necessarily are non-BPS orbits (because they cannot saturate any BPS bound)
and, in light of the above reasoning, they correspond to $\mathcal{N}=6$
orbits with $X=0$ representative. These simple arguments, combined with the
nilpotent orbits' analysis summarized in Table V of \cite{Bossard:2009at},
allows one to determine the intertwining supersymmetry-preserving properties
related to the charge orbits, listed in the Table below (we use the orbit
nomenclature reported in \autoref{tab:Mcharges}, and for small orbits the representatives are reported in brackets):
\begin{equation}\label{eq:N6}
\begin{array}{lllll}
\begin{array}{c}
\mathcal{O} \\
~%
\end{array}
& ~ &
\begin{array}{c}
\mathcal{N}=2,~J_{3}^{\mathbb{H}} \\
~%
\end{array}
& ~ &
\begin{array}{c}
\mathcal{N}=6,~J_{3}^{\mathbb{H}} \\
~%
\end{array}
\\
4a &  & 1/2\text{-BPS} &  & \text{nBPS}:\text{~}X_{H}\neq 0,Z_{AB,H}=0 \\
4b &  & \text{nBPS}:~Z_{H}=0 &  & 1/6\text{-BPS}:X_{H}=0,Z_{AB,H}\neq 0 \\
4c &  & \text{nBPS}:~Z_{H}\neq 0 &  & \text{nBPS}:~X_{H}\neq 0,Z_{AB,H}\neq 0
\\
3a &  & \text{nBPS~(}Z=0\text{)} &  & 1/6\text{-BPS~(}X=0\text{)} \\
3b &  & 1/2\text{-BPS~(}Z\neq 0\text{)} &  & \text{nBPS~(}X\neq 0\text{)} \\
2a &  & \text{nBPS~(}Z=0\text{)} &  & 1/3\text{-BPS~(}X=0\text{)} \\
2b &  & 1/2\text{-BPS~(}Z\neq 0\text{)} &  & 1/6\text{-BPS~(}X\neq 0\text{)}
\\
1 &  & 1/2\text{-BPS~(}Z\neq 0\text{)} &  & 1/2\text{-BPS~(}X\neq 0\text{)}%
\end{array}%
\end{equation}%
For analogue treatment in $D=5$, see \cite{Cerchiai:2010xv}.

\begin{sidewaystable}
\caption{Charge orbits, \textit{moduli spaces}, and number $\#$ of
"non-flat" scalar directions of the magic $D=4, \mathcal{N}=2$
supergravities defined over $\FA, \mathds{A}=\R, \C, \Q, \Oct$.
$M=\Aut(\FA)/mcs(\JA)$. $\dim_\R M=6+6\dim
\alg$ \cite{Ferrara:1997uz}.}\label{tab:Mcharges}
\begin{ruledtabular}\begin{tabular*}{\textwidth}{@{\extracolsep{\fill}}M{l}lM{l}|*{3}{M{c}}|*{3}{M{c}}}
\toprule \toprule \multirow{2}{*}{\textrm{Rank}}&\multirow{2}{*}{BH}
& \multirow{2}{*}{Susy}  & \multicolumn{3}{c}{$\FO$, $n_V=27$,
$M=E_{7(-25)}/[\U(1)\times E_{6(-78)}]$}  &
\multicolumn{3}{c}{$\FH$, $n_V=15$,  $M=\SO^\star(12)/\U(6)$}     \\
[8pt]
&&&\text{Orbit}~\mathcal{O}&\text{Moduli space}~\mathcal{M}&\#&\text{Orbit}~\mathcal{O}&\text{Moduli space}~\mathcal{M}&\#\\
\toprule
 1      &{\tiny small d. critical}  &1/2     &  \frac{E_{7(-25)}}{E_{6(-26)}\ltimes \R^{27}}        &  \frac{E_{6(-26)}}{F_{4(-52)}}\ltimes {\scriptstyle\R^{27}}&1         & \frac{\SO^\star(12)}{\SU^\star(6)\ltimes \R^{15}}                 &  \frac{\SU^\star(6)}{\Usp(6)}\ltimes {\scriptstyle\R^{15}  } &1         \\[8pt]
 2a     &{\tiny small critical  }   &0      &  \frac{E_{7(-25)}}{\SO(2,9)\ltimes \R^{32}\oplus\R}   &  \frac{\SO(2,9)}{\SO(2)\times\SO(9)}\ltimes {\scriptstyle\R^{32}\oplus\R}&3   & \frac{\SO^\star(12)}{ [\SO(2,5)\times\SO(3)]\ltimes \R^{(8,2)}\oplus\R}       &  \frac{\SO(2,5)}{\SO(2)\times\SO(5)}\ltimes{\scriptstyle \R^{8}\oplus\R^{8}\oplus\R }  &3         \\[8pt]
 2b     &{\tiny small critical}     &1/2        &  \frac{E_{7(-25)}}{\SO(1,10)\ltimes \R^{32}\oplus\R}  &  \frac{\SO(1,10)}{\SO(10)}\ltimes {\scriptstyle\R^{32}\oplus\R}&11        & \frac{\SO^\star(12)}{ [\SO(1,6)\times\SO(3)]\ltimes \R^{(8,2)}\oplus\R}       &  \frac{\SO(1,6)}{\SO(6)}\ltimes \R^{8}\oplus{\scriptstyle\R^{8}\oplus\R}   &7         \\[8pt]
 3a     &{\tiny small light-like}   &0          &  \frac{E_{7(-25)}}{F_{4(-20)}\ltimes \R^{26}}     &  \frac{F_{4(-20)}}{\SO(9)}\ltimes {\scriptstyle\R^{26}}&12            & \frac{\SO^\star(12)}{\Usp(2,4)  \ltimes \R^{14}}                  &  \frac{\Usp(2,4)}{\Usp(2)\times\Usp(4)}\ltimes {\scriptstyle\R^{14}}   &8         \\[8pt]
 3b     &{\tiny small light-like}   &1/2    &  \frac{E_{7(-25)}}{F_{4(-52)}\ltimes \R^{26}}     &  \R^{26}&28                           & \frac{\SO^\star(12)}{ \Usp(6)\ltimes \R^{14}}                     &  \R^{14}   &1 6        \\[8pt]
 4a     &{\tiny large time-like}    &1/2    &  \frac{E_{7(-25)}}{E_{6(-78)}}                &  -                        &54         & \frac{\SO^\star(12)}{\SU(6)}                              &  -   &30       \\[8pt]
 4b     &{\tiny large time-like}    &0 \;{\scriptstyle (Z_H=0)}         &  \frac{E_{7(-25)}}{ E_{6(-14)}}               &  \frac{E_{6(-14)}}{\SO(10)\times \SO(2)}      &22 & \frac{\SO^\star(12)}{\SU(4,2)}                            &  \frac{\SU(4,2)}{\SU(4)\times\SU(2)}   &13         \\[8pt]
 4c     &{\tiny large space-like}   &0  \;{\scriptstyle(Z_H\not=0)}     &  \frac{E_{7(-25)}}{E_{6(-26)}}                &  \frac{E_{6(-26)}}{F_{4(-52)}}&28             & \frac{\SO^\star(12)}{\SU^\star(6)}                        &  \frac{\SU^\star(6)}{\Usp(6)}   &16        \\[8pt]
\toprule \toprule \multirow{2}{*}{\textrm{Rank}}&\multirow{2}{*}{BH}
& \multirow{2}{*}{Susy}  & \multicolumn{3}{c}{$\FC$, $n_V=9$,
$M=\SU(3,3)/[\U(1)\times\SU(3)\times\SU(3)]$}  &
\multicolumn{3}{c}{$\FTSR$, $n_V=6$,  $M=\Sp(6,\R)/\U(3)$}     \\
[8pt]
&&&\text{Orbit}~\mathcal{O}&\text{Moduli space}~\mathcal{M}&\#&\text{Orbit}~\mathcal{O}&\text{Moduli space}~\mathcal{M}&\#\\
\toprule
 1      &{\tiny small d. critical}&1/2  &  \frac{\SU(3,3)}{\SL(3,\C)\ltimes \R^{9}}                  &  \frac{\SL(3,\C)}{\SU(3)}\ltimes {\scriptstyle\R^{9}}&1                       & \frac{\Sp(6,\R)}{\SL(3,\R)\ltimes \R^{6}}         &  \frac{\SL(3,\R)}{\SO(3)}\ltimes {\scriptstyle\R^{6}} &1         \\[8pt]
 2a     &{\tiny small critical  }   &0    &  \frac{\SU(3,3)}{[\SO(2,3)\times\SO(2)]\ltimes \R^{(4,2)}\oplus \R }    &  \frac{\SO(2,3)}{\SO(2)\times\SO(3)}\ltimes{\scriptstyle \R^{4}\oplus\R^{4}\oplus\R}&3    & \frac{\Sp(6,\R)}{\SO(2,2)\ltimes \R^{4}\oplus \R}     &  \frac{\SO(2,2)}{\SO(2)\times\SO(2)}\ltimes{\scriptstyle \R^{4}\oplus\R}   &3         \\[8pt]
 2b     &{\tiny small critical}     &1/2  &  \frac{\SU(3,3)}{ [\SO(1,4)\times\SO(2)]\ltimes \R^{(4,2)}\oplus \R}    &  \frac{\SO(1,4)}{\SO(4)}\ltimes {\scriptstyle\R^{4}\oplus\R^{4}\oplus\R}&5            & \frac{\Sp(6,\R)}{\SO(1,3)\ltimes \R^{4}\oplus \R }    &  \frac{\SO(1,3)}{\SO(3)}\ltimes {\scriptstyle\R^{4}\oplus\R }  &4         \\[8pt]
 3a     &{\tiny small light-like}   &0    &  \frac{\SU(3,3)}{\SU(1,2)\ltimes\R^8}                   &  \frac{\SU(1,2)}{\U(1)\times\SU(2)}\ltimes {\scriptstyle\R^{8}}&6                 & \frac{\Sp(6,\R)}{\SU(1,1)\ltimes\R^5}         &  \frac{\SU(1,1)}{\U(1)\times\U(1)}\ltimes{\scriptstyle \R^{5}}   &6         \\[8pt]
 3b     &{\tiny small light-like}   &1/2  &  \frac{\SU(3,3)}{\SU(3)\ltimes\R^8}                     &  \R^{8}&10                                                & \frac{\Sp(6,\R)}{\SU(2)\ltimes\R^5}           &  \R^{5}   &7       \\[8pt]
 4a     &{\tiny large time-like}    &1/2  &  \frac{\SU(3,3)}{\SU(3) \times \SU(3)} &  -                     &18                         & \frac{\Sp(6,\R)}{\SU(3)}                      &  -   &12       \\[8pt]
 4b     &{\tiny large time-like}    &0 \;{\scriptstyle(Z_H=0)}&\frac{\SU(3,3)}{\SU(1, 2)  \times\SU(1,2)}               &  \frac{\SU(1, 2)  \times\SU(1,2)}{[\U(1)\times \SU(2)]^2}     &9              & \frac{\Sp(6,\R)}{\SU(1, 2)}                   &  \frac{\SU(1, 2)}{\U(1)\times\SU(2)}   &8         \\[8pt]
 4c &{\tiny large space-like}&0\;{\scriptstyle(Z_H\not=0)}&\frac{\SU(3,3)}{\SL(3, \C)}                              &  \frac{\SL(3, \C)}{\SU(3)}&10                                 & \frac{\Sp(6,\R)}{\SL(3, \R)}                  &  \frac{\SL(3, \R)}{\SO(3)}   &7       \\[8pt]
\bottomrule
\bottomrule
\end{tabular*}\end{ruledtabular}
\end{sidewaystable}


\subsection{The $\mathcal{N}=4$ and $\mathcal{N}=2$ Reducible Jordan Symmetric Sequences\label{sec:N4N2}}

\subsubsection{$\mathcal{N}=4$}
For $\mathcal{N}=4$ supergravity coupled to $n_V$ vector multiplets,
the $(n+6)+(n+6)$ electric$+$magnetic BH charges (where
$n=n_V\geqslant 0$) may be represented as elements
\begin{equation}
x=\begin{pmatrix}-q_0&P\\Q&p^0\end{pmatrix}, \quad\text{where}
\quad p^0, q^0\in\R\quad\text{and}\quad Q, P\in\Jnf
\end{equation}
of the Freudenthal triple system
$\mathfrak{F}^{6,n}:=\mathfrak{F}(\Jnf)$.   The details may be found
in \autoref{sec:FTS} of \cite{Borsten:2011nq}, and in Refs. therein.
The \textit{reducible} $D=4$, $\mathcal{N}=4$ U-duality group is
given by the automorphism group $\Aut(\Fnf)=\text{Conf}\left(
\frak{J}_{5,n-1}\right)=\SL(2,\R)\times\SO(6, n)$ under which $x\in
\Fnf$ transforms as a $\rep{(2,6+n)}$. The BH entropy is once again
given by Eq. (\ref{BH-entropy-D=4}), where $I_4(x)=\Delta(x)=\half q(x)$
is the unique quartic invariant polynomial of $\SL\left(
2,\mathds{R}\right) \times \SO\left( 6,n\right)$. The U-duality
charge orbits are classified according to the $\SL(2,
\R)\times\SO(6,n)$-invariant FTS \textit{rank} of the charge vector.
More precisely, we have the following theorem \cite{Borsten:2011nq}.
\begin{theorem}\label{thm:N4} Every BH charge vector $x\in\Fnf$ of a given rank is $\SL(2, \R)\times\SO(6,n)$ related one of the following canonical forms:
\begin{enumerate}
\item Rank 1
\begin{enumerate}
\item  $x_{1}=\begin{pmatrix}1&0\\0&0\end{pmatrix}$
\end{enumerate}
\item Rank 2
\begin{enumerate}
\item  $x_{2a}=\begin{pmatrix}1&E_1\\0&0\end{pmatrix}$
\item  $x_{2b}=\begin{pmatrix}1&-E_1\\0&0\end{pmatrix}$
\item  $x_{2c}=\begin{pmatrix}1&E_2\\0&0\end{pmatrix}$
\end{enumerate}
\item Rank 3
\begin{enumerate}
\item  $x_{3a}=\begin{pmatrix}1&E_2+E_3\\0&0\end{pmatrix}$
\item  $x_{3b}=\begin{pmatrix}1&E_2-E_3\\0&0\end{pmatrix}$
\end{enumerate}
\item Rank 4
\begin{enumerate}
\item   $x_{4a}=k\begin{pmatrix}1&-E_1+E_2+E_3\\0&0\end{pmatrix}$
\item   $x_{4b}=k\begin{pmatrix}1&E_1+E_2-E_3\\0&0\end{pmatrix}$
\item   $x_{4c}=k\begin{pmatrix}1&-E_1+E_2-E_3\\0&0\end{pmatrix}$
\end{enumerate}
\end{enumerate}
where $k>0$ and the $E_i$ are as given in \eqref{eq:N4idem}.
\end{theorem}

The orbit stabilizers are summarized in \autoref{tab:n4redorbits}.

\begin{sidewaystable}
\caption{Charge orbits, \textit{moduli spaces}, the number $\#$ of
``non-flat" scalar directions of the reducible $D=4, \mathcal{N}=4$
supergravities defined over
$\FTS^{6,n}:=\FTS(\mathfrak{J}_{5,n-1})$. $M=[\SL(2,
\R)\times\SO(6,n)]/[\SO(2)\times\SO(6)\times\SO(n)]$.
$\dim_\R(M)=6n+2$. For comparison we have included the orbit
labeling used in \cite{Cerchiai:2009pi}, and then in
\cite{Andrianopoli:2010bj} and \cite{Ceresole:2010nm}.  The table is
split according as the BHs are small or
large.}\label{tab:n4redorbits}
\begin{ruledtabular}\begin{tabular*}{\textwidth}{@{\extracolsep{\fill}}M{l}lM{l}M{c}M{c}M{c}M{l}}
\toprule
\toprule
 \text{{\normalsize Rank}}      & {\normalsize BH}                  &\text{{\normalsize Susy}}              & {\normalsize\text{Charge orbit $\mathcal{O}$}}                                                                & {\normalsize\text{Moduli space $\mathcal{M}$}}    & \#    \\
\toprule
    1/\text{\textbf{A.3}}   &  {d. critical}  &1/2                &\frac{\SL(2, \R)\times\SO(6,n)}{[\SO(1,1)\times \SO(5,n-1)]\ltimes (\R\times \R^{5,n-1})}          &\frac{\SO(1,1)\times \SO(5,n-1)}{\SO(5)\times\SO(n-1)}\ltimes {\scriptstyle \R\times \R^{5,n-1}} & 1\\[12pt]
    2a/\text{\textbf{A.2}}  &  {critical}     &0                  &\frac{\SL(2, \R)\times\SO(6,n)}{\SO(6,n-1)\times \R }                                          &\frac{\SO(6,n-1)}{\SO(6)\times\SO(n-1)}\ltimes {\scriptstyle \R} & 7\\[12pt]
    2b/\text{\textbf{A.1}}  &  {critical}     &1/2                &\frac{\SL(2, \R)\times\SO(6,n)}{\SO(5,n)\times \R}                                         &\frac{\SO(5,n)}{\SO(5)\times\SO(n)}\ltimes {\scriptstyle \R}&2n+2\\[12pt]
    2c/\text{\textbf{B} }   &  {critical}     &1/4                &\frac{\SL(2, \R)\times\SO(6,n)}{[\SO(2,1)\ltimes \R]\times[\SO(4,n-2) \ltimes (\R^{4,n-2}\oplus\R^{4,n-2})]}&\frac{\SO(2,1)\times\SO(4,n-2)}{\SO(2)\times\SO(4)\times\SO(n-2)}\ltimes {\scriptstyle \R\times [\R^{4,n-2}\oplus\R^{4,n-2}]}&4 \\        [12pt]
    3a/\text{\textbf{C.1}}  &  {light-like}   &1/4                &\frac{\SL(2, \R)\times\SO(6,n)}{[\SO(4,n-1)\ltimes \R^{4,n-1}]\times\R}                                &\frac{\SO(4,n-1)}{\SO(4)\times\SO(n-1)}\ltimes{\scriptstyle \R\times \R^{4, n-1}}& n\\[12pt]
    3b/\text{\textbf{C.2}}  &  {light-like}   &0                  &\frac{\SL(2, \R)\times\SO(6,n)}{[\SO(5,n-2)\ltimes \R^{5,n-2}]\times\R}                            &\frac{\SO(5,n-2)}{\SO(5)\times\SO(n-2)}\ltimes{\scriptstyle \R\times \R^{5, n-2}}&8\\[12pt]
\toprule
    4a/\alpha           &  {time-like}        &1/4                &\frac{\SL(2, \R)\times\SO(6,n)}{\SO(2)\times \SO(4,n)}                                             &\frac{\SO(4,n)}{\SO(4)\times\SO(n)}&2n+2\\[12pt]
    4b/\gamma       &  {time-like}        &0\; {(\hat{Z}_{AB,H}=0)}  &\frac{\SL(2, \R)\times\SO(6,n)}{\SO(2)\times \SO(6,n-2)}                                       &\frac{\SO(6, n-2)}{\SO(6)\times\SO(n-2)}&14\\[12pt]
    4c/\beta            &  {space-like}   &0\; {(\hat{Z}_{AB,H}\not=0)} &\frac{\SL(2, \R)\times\SO(6,n)}{\SO(1,1)\times \SO(5,n-1)}                              &\frac{\SO(1,1)\times \SO(5,n-1)}{\SO(5)\times\SO(n-1)}&n+6\\[12pt]
\bottomrule
\bottomrule
\end{tabular*}\end{ruledtabular}
\end{sidewaystable}

\subsubsection{$\mathcal{N}=2$}
For $\mathcal{N}=2$ supergravity theories coupled to $n_V$ vector
multiplets whose scalar manifolds belong to the so-called Jordan
symmetric sequence of special K\"{a}hler geometry, the $(n+2)+(n+2)$
electric$+$magnetic BH charges (where $n=n_V-1\geqslant 1$) may be
represented as elements
\begin{equation}
x=\begin{pmatrix}-q_0&P\\Q&p^0\end{pmatrix}, \quad\text{where}
\quad p^0, q^0\in\R\quad\text{and}\quad Q, P\in\Jnt
\end{equation}
of the Freudenthal triple system
$\mathfrak{F}^{2,n}:=\mathfrak{F}(\Jnt)$.   The details may be found
in \autoref{sec:FTS} of \cite{Borsten:2011nq}, as well as in Refs.
therein. The \textit{reducible} $D=4$, $\mathcal{N}=2$ U-duality
group is given by the automorphism group
$\Aut(\Fnt)\cong\text{Conf}\left(
\frak{J}_{1,n-1}\right)=\SL(2,\R)\times\SO(2, n)$ under which $x\in
\Fnt$ transforms as a $\rep{(2,2+n)}$. The BH entropy is once again
given by Eq. (\ref{BH-entropy-D=4}), where $I_4(x)=\Delta(x)=\half q(x)$
is the unique quartic invariant polynomial of $\SL\left(
2,\mathds{R}\right) \times \SO\left( 2,n\right)$. The U-duality
charge orbits are classified according to the $\SL(2,\R)\times\SO(2,
n)$-invariant FTS \textit{rank} of the charge vector. The orbit
representatives are as in \autoref{thm:N4} \cite{Borsten:2011nq}.
However, physically each 1/4-BPS orbits of
\autoref{tab:n4redorbits} splits into one 1/2-BPS orbit and one
non-BPS orbit, see \autoref{tab:n2redorbits}. This splitting is
determined by the sign of the quantity \cite{Bellucci:2006xz}
\begin{equation}
\mathcal{I}_2=|Z|^2-|D_S Z|^2 \label{I2-quantity}.
\end{equation}
Here, $Z$ is the central charge  and $D_S Z$ is the axion-dilaton
\textit{matter charge}, where $D_S$  is the K\"{a}hler covariant
derivative on the scalar manifold along the axion-dilaton direction;
this is a ``privileged" scalar direction, because the scalar manifold
is factorized. In fact, noting that the $\mathcal{N}=4$, $D=4$
1/4-BPS canonical forms all have a Jordan algebra element that has
two disconnected components under $\Str_0(\Jnt)$, the sign condition
on (\ref{I2-quantity}) can be rephrased in terms of the charges.
\begin{sidewaystable}
\caption{Charge orbits, \textit{moduli spaces}, and number $\#$ of
``non-flat" scalar directions of the reducible $D=4, \mathcal{N}=2$
supergravities defined over
$\FTS^{2,n}:=\FTS(\mathfrak{J}_{1,n-1})$. $M=[\SL(2,
\R)\times\SO(2,n)]/[\SO(2)^2\times\SO(n)]$. $\dim_\R(M)=2n+2$. For
comparison, we have included the orbit labelling used in
\cite{Cerchiai:2009pi}, and then in \cite{Andrianopoli:2010bj} and
\cite{Ceresole:2010nm}. The table is split according as the BHs are
small or large.}\label{tab:n2redorbits}
\begin{ruledtabular}\begin{tabular*}{\textwidth}{@{\extracolsep{\fill}}M{l}clM{l}M{c}M{c}M{c}M{l}}
\toprule
\toprule
 \text{Rank} & \cite{Cerchiai:2009pi}       & BH                    &\text{Susy}                & \text{Charge orbit}~ \mathcal{O}                                                              & \text{Moduli space}~ \mathcal{M}  & \#    \\
\toprule
    1                   &   {\textbf{A.3}}  &  {d. critical} &1/2                &\frac{\SL(2, \R)\times\SO(2,n)}{[\SO(1,1)\times \SO(1,n-1)]\ltimes (\R\times \R^{1, n-1})}         &\frac{\SO(1,1)\times \SO(1,n-1)}{\SO(n-1)}\ltimes {\R\times \R^{1, n-1}} & 1\\[12pt]
    2a                  &   {\textbf{A.2}}  &  {critical}        &0                  &\frac{\SL(2, \R)\times\SO(2,n)}{\SO(2,n-1)\times \R }                                          &\frac{\SO(2,n-1)}{\SO(2)\times\SO(n-1)}\ltimes {\R} & 3 \\[12pt]
    2b                  &   {\textbf{A.1}}  &  {critical}        &1/2                &\frac{\SL(2, \R)\times\SO(2,n)}{\SO(1,n)\times \R }                                            &\frac{\SO(1,n)}{\SO(n)}\ltimes {\R}& n+1\\[12pt]
    2c^{+}              &   {\textbf{B}}        &  {critical}        &1/2\; {\mathcal{I}_2>0}                   &\frac{\SL(2, \R)\times\SO(2,n)}{[\SO(2,1)\ltimes\R]\times[\SO(n-2)\ltimes (\R^{n-2}\oplus\R^{n-2})]}&\frac{\SO(2,1)}{\SO(2)}\ltimes {\R\times [\R^{n-2}\oplus\R^{n-2}}]& 3 \\     [12pt]
    2c^{-}              &   {\textbf{B}}        &  {critical}        &0\; {\mathcal{I}_2<0}                 &\frac{\SL(2, \R)\times\SO(2,n)}{[\SO(2,1)\ltimes\R]\times[\SO(n-2)\ltimes (\R^{n-2}\oplus\R^{n-2})]}&\frac{\SO(2,1)}{\SO(2)}\ltimes {\R\times [\R^{n-2}\oplus\R^{n-2}}]& 3\\      [12pt]
    3a^{+}              &   {\textbf{C.1}}  &  {light-like}  &1/2\; {\mathcal{I}_2>0}               &\frac{\SL(2, \R)\times\SO(2,n)}{[\SO(n-1)\ltimes \R^{n-1}]\times\R}                                &{\R\times \R^{n-1}}& n+2\\[12pt]
    3a^{-}              &   {\textbf{C.1}}  &  {light-like}  &0\; {\mathcal{I}_2<0}             &\frac{\SL(2, \R)\times\SO(2,n)}{[\SO(n-1)\ltimes \R^{n-1}]\times\R}                                &{\R\times \R^{n-1}}& n+2\\[12pt]
    3b                  &   {\textbf{C.2}}  &  {light-like}  &0                  &\frac{\SL(2, \R)\times\SO(2,n)}{[\SO(1,n-2)\ltimes \R^{n-1}]\times\R}                          &\frac{\SO(1,n-2)}{\SO(n-2)}\ltimes {\R^{n-1}\times\R}&4\\[12pt]
\toprule
    4a^{+}              & $\alpha$          &  {time-like}       &1/2\; {\mathcal{I}_2>0}               &\frac{\SL(2, \R)\times\SO(2,n)}{\SO(2)\times \SO(n)}                                           &-&2n+2\\[12pt]
    4a^{-}              & $\alpha$          &  {time-like}       &0\; {\mathcal{I}_2<0}             &\frac{\SL(2, \R)\times\SO(2,n)}{\SO(2)\times \SO(n)}                                           &-&2n+2\\[12pt]
    4b                  &   $\gamma$        &  {time-like}       &0\; {Z_H=0}   &\frac{\SL(2, \R)\times\SO(2,n)}{\SO(2)\times \SO(2,n-2)}                                       &\frac{\SO(2, n-2)}{\SO(2)\times\SO(n-2)}&8\\[12pt]
    4c                  & $\beta$           &  {space-like}  &0\; {Z_H\not=0} &\frac{\SL(2, \R)\times\SO(2,n)}{\SO(1,1)\times \SO(1,n-1)}                               &\frac{\SO(1,1)\times \SO(1,n-1)}{\SO(n-1)}&n+2\\[12pt]
\bottomrule
\bottomrule
\end{tabular*}\end{ruledtabular}
\end{sidewaystable}

\subsection{Interpretation of $\sharp
_{\frac{1}{2}-BPS,\text{rank-}1}=1$}

As reported in the Tables, all \textit{symmetric} $D=4$ theories
share the same result, namely:
\begin{equation}
\sharp _{\frac{1}{2}-BPS,\text{rank-}1}=1. \label{ress}
\end{equation}
Note that the rank-$1$, doubly critical orbit is always unique,
corresponding to the maximum weight vector in the relevant
representation space. Up to U-duality all  rank-1
$D=4$ black holes  may be regarded as a pure KK state of the 5-dimensional parent theory. All along the $\frac{1}{2}$-BPS rank-$1$
scalar flow \cite {Andrianopoli:2010bj}, there is only one
``non-flat'' scalar degree of freedom.

This can be easily interpreted by recalling that the first-order
superpotential of the $\mathcal{N}=2$ BPS flows is nothing but $\mathcal{W}%
=\left| Z\right| $, where $Z$ is the $\mathcal{N}=2$ central charge
\cite {Ceresole:2007wx}. Thus, by considering the general expression
of $Z$ in a generic $d$-special K\"{a}hler geometry (given by Eq.
(4.9) of \cite {Ceresole:2007rq}) for the relevant representative
$1$-charge configuration in which the dependence on only one scalar
field is manifest (which turns out to be $\left\{ q_{0}\right\} $),
one obtains:
\begin{equation}
\mathcal{W}_{\frac{1}{2}-BPS,\text{rank-}1}=\left| Z\right|
_{\left\{
q_{0}\right\} }=\frac{\left| q_{0}\right| }{2\sqrt{2}}\mathcal{V}^{-1/2}%
\text{,}
\end{equation}
where $\mathcal{V}\equiv r_{KK}^{3}$, $r_{KK}$ denoting the KK
radius in the KK reduction $D=5\longrightarrow D=4$
\cite{Ceresole:2007rq}.

In the cases $\mathcal{N}=8$ and $\mathcal{N}=4$, similar results
can be obtained from the treatment given in \cite{Ceresole:2009id}
and \cite {Cerchiai:2009pi}. Analogous explanations can be given for
the result (\ref {ress}) for $D=5$ charge orbits, as reported in the
relevant Tables.

\subsection{The  $\mathcal{N}=2$ $STU$, $ST^2$ and $T^3$ Models\label{sec:degen}}

\subsubsection{$STU$}\label{sec:STU}

The $STU$ model is $\mathcal{N}=2$ supergravity coupled to  three
vector multiplets. However, it has an additional discrete triality,
which exchanges the roles of the three complex moduli. This triality
has a stringy explanation first identified in \cite{Duff:1995sm}. It
is essentially  a remnant of the  $D=6$  equivalence between the
heterotic string on $T^4$, the Type IIA string on $K3$ and the Type
IIB string on its mirror. The $STU$ model is thus a noteworthy
element ($n=2$) of the $\mathcal{N}=2$, $D=4$ Jordan symmetric
sequence discussed above.

The $(1+3)+(1+3)$ electromagnetic charges may be represented as
elements
\begin{equation}
x=\begin{pmatrix}-q_0&(p; p^\mu)\\(q; q_\nu)&p^0\end{pmatrix}, \quad\text{where}
\quad p^0, q^0\in\R\quad\text{and}\quad (q; q_\nu), (p; p^\mu) \in\J_{1,1}
\end{equation}
of the Freudenthal triple system $\mathfrak{F}^{2,2}:=\mathfrak{F}(\J_{1,1})$.\\
The U-duality group $\Aut(\FTS_{STU})\cong\text{Conf}\left( \frak{J}_{1,1}=\mathds{R}\oplus \mathbf{\Gamma }_{1,1}=%
\mathds{R}\oplus \mathds{R}\oplus \mathds{R}\right)=\SL(2, \R)\times
\SO(2,2)$ may be recast in a form reflecting this triality symmetry
using the isomorphism $\SO(2,2)\cong\SL(2, \R)\times \SL(2, \R)$.
From the the heterotic string perspective this corresponds to an
$\SL(2, \Z)_S$ strong/weak coupling duality and an $\SL(2, \Z)_T
\times \SL(2, \Z)_U$ target space duality acting on the
dilaton/axion, complex K\"{a}hler form and the complex structure
fields $S, T, U$ respectively. At the level of the FTS
\cite{Borsten:2008, Borsten:2008wd, Borsten:2009yb}, this is
realised by the Jordan algebra isomorphism $\J_{1,1}=\R\oplus
\Gamma_{1,1}\cong\R\oplus\R\oplus\R=\J_{STU}$ which, for $(q_1, q_2,
q_3)\in\J_{STU}$ and $(q; q_\nu)\in\J_{1,1}$ is given by, \be q_1=q,
\quad q_2=q_0+q_1, \quad q_3=q_0-q_1, \ee so that the $STU$ cubic
norm becomes \be N(Q)=q_1q_2q_3. \ee By renaming \be
\begin{pmatrix}-q_0&(p_1, p_2, p_3)\\(q_1, q_2, q_3)&p^0\end{pmatrix}\mapsto \begin{pmatrix}a_{000}&(a_{011}, a_{101}, a_{110})\\(a_{100}, a_{010},
a_{001})&a_{111}\end{pmatrix}, \ee the charges may be arranged into
a $2\times 2 \times 2$ hypermatrix $a_{ABC}$, where $A, B, C=0, 1$,
transform as a $(\rep{2,2,2})$ under $\SL_A(2, \R)\times \SL_B(2,
\R)\times \SL_C(2, \R)$. In such a way, the quartic norm is given by
Cayley's hyperdeterminant $\Det a_{ABC}$ \cite{Cayley:1845,
Duff:2006uz}, \be \Delta=-\Det
a=\frac{1}{2}\epsilon^{A_1A_2}\epsilon^{B_1B_2}\epsilon^{C_1C_3}\epsilon^{A_3A_4}\epsilon^{B_3B_4}\epsilon^{C_2C_4}a_{A_1B_1C_1}a_{A_2B_2C_2}a_{A_3B_3C_3}a_{A_4B_4C_4}
\ee and \be S_{D=4, \text{BH}}=\pi\sqrt{|\Det a|}.
\label{entropy-stu} \ee This observation lies at the origin of the
``black-hole/qubit correspondence''
\cite{Levay:2006kf,Duff:2006ue,Levay:2006pt,Duff:2007wa,Levay:2007nm,Borsten:2008ur,Borsten:2008,Levay:2008mi,Borsten:2008wd,Levay:2009bp,Borsten:2009ae,Levay:2010qp,Levay:2010ua,
Borsten:2010db,Borsten:2011is,Rios:2011fa}. The hyperdeterminant is manifestly
invariant under the triality $A\leftrightarrow B\leftrightarrow C$.
The role of more general hyperdeterminants in M-theory can be found
in \cite{Fang:2010vs, Gibbs:2010uz}.

The implication of this triality for the structure of the orbits is
that what are distinct cosets for generic $n_V$ become isomorphic
for the $STU$  case. In particular, we find that for the $STU$ model
\cite{Borsten:2009yb} \be
\mathcal{O}_{2a}\cong\mathcal{O}_{2b}\cong\mathcal{O}_{2c}, \qquad
\mathcal{O}_{3a}\cong\mathcal{O}_{3b} \ee as can be seen immediately
from \autoref{tab:n2redorbits} setting $n=2$. However, while the
cosets are isomorphic the distinct physical properties of each orbit
are preserved, so that the $STU$ model can really be included in the
generic sequence.

\subsubsection{$ST^2$}\label{sec:ST2}

On the other hand, the orbit structure of the $ST^2$ model, which
can be seen as the first ($n=1$) element of the Jordan symmetric
sequence, $\mathcal{N}=2$ coupled to two vector multiplets, does
depart from the one discussed so far. The $(1+2)+(1+2)$
electromagnetic charges may be represented as elements
\begin{equation}
x=\begin{pmatrix}-q_0&(p^1, p^2)\\(q_1, q_2)&p^0\end{pmatrix},
\quad\text{where} \quad p^0, q^0\in\R\quad\text{and}\quad (p^1,
p^2), (q_1, q_2) \in\R\oplus\R
\end{equation}
of the Freudenthal triple system $\mathfrak{F}^{2,1}:=\mathfrak{F}(\J_{1})$.  Here, $\J_{1}=\R\oplus\Gamma_1=\R\oplus\R$ now has an ``Euclidean'' cubic norm
\be
N(Q)=q_1(q_{2})^{2}, \quad Q\in\J_{ST^2},
\ee
which implies there is only one rank 2 $Q\in\J_{ST^2}$ up to $\Str_0(\J_{ST^2})=\SO(1,1)$, which is now pure dilatation.  Consequently, the third rank 2 orbit (in the FTS) of the generic sequence $(n_V\geq3)$ vanishes \cite{Borsten:2011nq}.

The U-duality group is $\Aut(\FTS_{ST^2})\cong\text{Conf}\left(
\mathds{R}\oplus \mathds{R}\right)=\SL_A(2, \R)\times \SL_B(2,\R)$
under which the charges transform as a $(\rep{2,3})$. Again, this
symmetry is made manifest by writing the charges as a hypermatrix
\be Q=a_{A(B_1B_2)}. \ee The BH entropy is given by Eq.
(\ref{entropy-stu}), with the hyperdeterminant now being the ``$ST^2$
degeneration" of the expression holding for the $STU$ model (see
e.g. \cite{Bellucci:2007zi} for further details). The canonical
forms are presented in \autoref{thm:st2canform} \cite{Borsten:2011nq}. The orbits may be
obtained from \autoref{tab:n2redorbits}  by setting $n=1$ (when this
is still well defined - when it is not, the orbit is not present).

\begin{theorem}\label{thm:st2canform} \emph{\cite{Borsten:2011nq}} Every element $x\in \FTS_{ST^2}$ of a given rank is $\SL(2, \R)\times\SL(2, \R)$ related to one of the following canonical forms:
\begin{enumerate}
\item Rank 1
\begin{enumerate}
\item  $x_{1}=\begin{pmatrix}1&0\\0&0\end{pmatrix}$
\end{enumerate}
\item Rank 2
\begin{enumerate}
\item  $x_{2a}=\begin{pmatrix}1&(1;0)\\0&0\end{pmatrix}$
\item  $x_{2b}=\begin{pmatrix}1&(-1;0)\\0&0\end{pmatrix}$
\end{enumerate}
\item Rank 3
\begin{enumerate}
\item  $x_{3a}=\begin{pmatrix}1&(0;1)\\0&0\end{pmatrix}$
\end{enumerate}
\item Rank 4
\begin{enumerate}
\item   $x_{4a}=k\begin{pmatrix}1& (-1;1)\\0&0\end{pmatrix}$
\item   $x_{4b}=k\begin{pmatrix}1& (1;1)\\0&0\end{pmatrix}$
\end{enumerate}
\end{enumerate}
\end{theorem}

\subsubsection{$T^3$}\label{sec:T3}

Finally, we come to the $T^3$ model. Unlike all the other cases
treated here, the $T^3$ has a cubic Jordan algebra,
$\J_{T^3}=\R$, with a single non-zero rank. The cubic norm is given by \be N(Q)=q^3, \quad q\in
\R. \ee Hence, there is \textit{only a single} rank given by $N(Q)\not=0$: all non-zero elements are rank 3. Consequently, the rank 2, where we now mean in the FTS $\FTS(\J_{T^3})$,
orbit disappears entirely \cite{Borsten:2011nq}. That is, if a small
BH is critical, then it is doubly critical.

The U-duality group is $\Aut(\FTS_{T^3})\cong\text{Conf}\left(
\mathds{R}\right)=\SL_A(2, \R)$ under which the charges transform as
a $\rep{4}$ (spin $s=3/2$).  Again, this symmetry is made manifest
by writing the charges as a hypermatrix \be Q=a_{(A_1A_1A_2)}. \ee
The BH entropy is given by Eq. (\ref{entropy-stu}), with the
hyperdeterminant now being the ``$T^3$ degeneration" of the
expression holding for the $STU$ model (see e.g.
\cite{Bellucci:2007zi} for further details). 

Accounting for the
vanishing rank 2 case, the remaining $\SL_A(2, \R)$-orbits are given
in \autoref{thm:T3}.  \begin{theorem}\label{thm:T3} \emph{\cite{Borsten:2011nq}} Every element $x\in \FTS_{T^3}$ of s given rank is $\SL(2, \R)$ related to one of the following canonical forms:
\begin{enumerate}
\item Rank 1
\begin{enumerate}
\item  $x_{1}=\begin{pmatrix}1&0\\0&0\end{pmatrix}$
\end{enumerate}
\item Rank 3
\begin{enumerate}
\item  $x_{3a}=\begin{pmatrix}0&1\\0&0\end{pmatrix}$
\end{enumerate}
\item Rank 4
\begin{enumerate}
\item   $x_{4a}=k\begin{pmatrix}1& -1\\0&0\end{pmatrix}$
\item   $x_{4b}=k\begin{pmatrix}1& 1\\0&0\end{pmatrix}$
\end{enumerate}
\end{enumerate}
\end{theorem}There are now just four
orbits: small doubly critical (rank 1) 1/2-BPS, small light-like
(rank 3) 1/2-BPS, large (rank 4) 1/2-BPS and non-BPS. This is
consistent with the analysis of \cite{Kim:2010bf,Fre:2011uy,Fre:2011aa}, which
relies on the theory of nilpotent orbits.  The BPS nature of both
``small'' (rank $3$ and rank $1$) charge orbits of this model can
also be easily understood by recalling the result derived in Sec.
5.5 of \cite{Andrianopoli:2010bj}, namely that the ``small'' limit
of the first-order (``fake'') superpotentials of both BPS and
non-BPS attractor scalar flows yields nothing but the absolute value
$\left| Z\right| $ of the $\mathcal{N}=2$ central charge.

Performing a time-like reduction (since we are interested in
stationary solutions) the resulting 3-dimensional $T^3$ model has
$G_{2(2)}$ U-duality, with scalars parametrising the
pseudo-Riemannian coset, \be \frac{G_{2(2)}}{\SO_0(2,2)}. \ee
 The nilpotent $\SO_0(2,2)$-orbits of $\mathfrak{g}_{2(2)}$ correspond to six static  (\emph{i.e.} single or non-interacting centre) extremal solutions \cite{Kim:2010bf}. However,  only four of these orbits, labeled $\mathcal{O}_1, \mathcal{O}_2, \mathcal{O}_{3K}, \mathcal{O}_{4K'}$ in \cite{Kim:2010bf}, correspond to physically acceptable static solutions \cite{Kim:2010bf}.  From our perspective the unphysical orbits cannot be seen and it can be checked that the four orbits we describe correspond precisely to the four physical orbits of \cite{Kim:2010bf,Fre:2011uy, Fre:2011aa}. Explicitly, where we use the labeling in \autoref{thm:T3},
\be
\begin{array}{lcll}
\mathcal{O}_{1} &\longleftrightarrow &\mathcal{O}_{x_{1}}& \text{small doubly critical (rank 1) 1/2-BPS,}\\
\mathcal{O}_2&\longleftrightarrow&\mathcal{O}_{x_{3}}& \text{small light-like (rank 3) 1/2-BPS,}\\
 \mathcal{O}_{3K}&\longleftrightarrow&\mathcal{O}_{x_{4a}}& \text{large (rank 4) 1/2-BPS,}\\
 \mathcal{O}_{4K'}&\longleftrightarrow&\mathcal{O}_{x_{4b}}& \text{large (rank 4) non-BPS.}
\end{array}
\ee The orbit stabilizers are summarized in \autoref{tab:orbitsT3}.
Note, the two large (1/2-BPS and non-BPS) orbits have no continuous
stabilizers. However, the 1/2-BPS case does have a discrete $\Z_3$
stabilizer generated by \be\label{eq:Z3}
M=\frac{1}{2}\begin{pmatrix}-1&\sqrt{3}\\-\sqrt{3}&-1\end{pmatrix},
\ee where $M\in\SL(2, \R)$. Note, this is a finite subgroup of the $\SL(2,\R)$ U-duality and should not be misconstrued as a sub-group the $STU$ triality symmetry, which collapses upon  identifying the moduli. The origin of $\Z_3$ is easily understood in terms of the ``parent'' 1/2-BPS rank-4  $STU$ orbit stabilizer $\SO(2)\times\SO(2)$. Recall, the Lie algebra of the automorphism group $\mathfrak{Aut(F(J))}$ decomposes under the reduced structure group $\Str_0(\J)$ according as
\be
\mathfrak{Aut(F(J))}=\mathfrak{Str}_0(\J)\oplus\J\oplus\J\oplus\R.
\ee 
The 1/2-BPS rank-4  $STU$ stability group is conjugate to\footnote{In fact, for our orbit representative, equal to.} an $\SO(2)\times\SO(2)$ generated by (using the notation introduced in \hyperref[sec:app1]{appendix A}) $\Phi=(0,X,-X,0)$,  $\Phi\in\mathfrak{Str}_0(\J)\oplus\J\oplus\J\oplus\R$, such that $\Tr(X)=0$. One possible parametrization of $\SO(2)\times\SO(2)\subset\SL_A(2,\R)\times\SL_B(2,\R)\times\SL_C(2,\R)$, obtained by exponentiating $\Phi$, is given by,
\be
\begin{pmatrix}\cos(\phi)&-\sin(\phi)\\\sin(\phi)&\cos(\phi)\end{pmatrix}\otimes\begin{pmatrix}\cos(\psi)&-\sin(\psi)\\\sin(\psi)&\cos(\psi)\end{pmatrix}\otimes\begin{pmatrix}\cos(\phi+\psi)&\sin(\phi+\psi)\\-\sin(\phi+\psi)&\cos(\phi+\psi)\end{pmatrix}.
\ee
Symmetrizing down from the $STU$ model to the $T^3$ model implies identifying the three factors appearing in the above parametrization. This gives \eqref{eq:Z3} and its powers, hence picking out a $\Z_3$ finite subgroup. Alternatively, this may be checked directly using the  totally symmetrized  hypermatrix, which transforms as
\be
a_{(A_1A_2A_3)}\mapsto \tilde{a}_{(A_1A_2A_3)}=M_{A_1}{}^{A'_1}M_{A_2}{}^{A'_2}M_{A_3}{}^{A'_3}a_{(A'_1A'_2A'_3)},
\ee
under $\SL(2, \R)$. Solving $\tilde{a}^{4a}_{(A_1A_2A_3)}=a^{4a}_{(A_1A_2A_3)}$, where $a^{4a}_{(A_1A_2A_3)}$ is the orbit representative appearing in \autoref{thm:T3},  yields the same conclusion. Since this $\Z_3$ forms a finite sub-group of a \emph{compact} stabilizer there should be no corresponding ``discrete'' moduli space.

By considering its embedding in the $STU$ model it is also particularly easy to see why there is no discrete stabilizer in the unique $\Delta<0$ non-BPS orbit. The  $\Delta<0$ non-BPS $STU$ orbit stabilizer  is conjugate to an $\SO(1,1)\times\SO(1,1)$ generated by $\Phi=(\phi,0,0,0)$,  $\phi\in\mathfrak{Str}_0(\J)$. Equivalently, there is a U-duality frame in which only the two graviphoton charges are turned on. Since the graviphotons are singlets under the $D=5$ U-duality group the stabilizer is precisely ${\Str}_0(\J)$. This is true for all $D=4$ theories  based on cubic Jordan algebras, explaining this common feature of the $\Delta<0$ non-BPS orbits. However, for the $T^3$ model ${\Str}_0(\J)$ contains only the identity, hence there can be no discrete stabilizer. This expectation is borne out by explicit computation. Note, since the presence of only graviphoton charges implies $\Delta<0$, this charge configuration is only possible for $\Delta<0$ non-BPS states.

\begin{table}[!ht]
\caption{Charge orbits, \textit{moduli spaces}, and number $\#$ of
"non-flat" scalar directions of the  $D=4, T^3$ model. $M=\SL(2,
\R)/\SO(2)$, $\dim_\R=2$. $L_{+}$ is the generator of
$\SL(2,\mathds{R})$ with positive grading with respect to its maximal
subgroup $\SO\left( 1,1\right) $.}\label{tab:orbitsT3}
\begin{ruledtabular}\begin{tabular*}{\textwidth}{@{\extracolsep{\fill}}M{l}cM{l}M{c}M{c}M{c}}
\toprule
 \toprule
\text{ Rank}&      BH               &\text{Susy}    & \text{Charge orbit}~ \mathcal{O}      & \text{Moduli space}~ \mathcal{M}  & \#    \\
\toprule
 1      &doubly critical    &1/2    & \frac{\SL(2,\R)}{L_+}              & \R    &   1       \\[8pt]

 3      &light-like &1/2    & \frac{\SL(2,\R)}{\mathds{1}}                  &   -   &   2   \\[8pt]
 \toprule
4 (\Delta>0)    &\multirow{2}{*}{large}     &1/2    & \frac{\SL(2,\R)}{\Z_3}                                  &-    &       2       \\[8pt]
 4 (\Delta<0)   &                   &0  & \frac{\SL(2,\R)}{\mathds{1}}              & -     &   2           \\[8pt]
\bottomrule
\bottomrule
\end{tabular*}\end{ruledtabular}
\end{table}

\subsection{\label{N=2-d=4-quadratic}$\mathcal{N}=2$
\textit{Minimally Coupled}}\label{sec:N2mc}

We now consider $\mathcal{N}=2$, $d=4$ ungauged supergravity \textit{%
minimally coupled} \textit{(mc)} \cite{Luciani:1977hp} to $n_{V}$
Abelian vector multiplets, whose scalar manifold is given by the
sequence of homogeneous symmetric \textit{rank-}$1$ special
K\"{a}hler manifolds
\begin{equation}
\mathcal{M}_{\mathcal{N}=2,mc,n}=\mathds{CP}^{n}\equiv \frac{G_{\mathcal{N}%
=2,mc,n}}{H_{\mathcal{N}=2,mc,n}}=\frac{\U(1,n)}{\U(n)\times \U(1)},~\text{dim}%
_{\mathds{R}}=2n,~n=n_{V}\in \mathds{N}.
\end{equation}
This theory cannot be uplifted to $D=5$, and it does not enjoy an
interpretation in terms of Jordan algebras. The $1+n$ vector field
strengths
and their duals, as well as their asymptotical fluxes, sit in the \textit{%
fundamental }$\mathbf{1+n}$ \ representation of the $\U$-duality group $G_{%
\mathcal{N}=2,mc,n}=\U\left( 1,n\right) $, in turn embedded in the
symplectic group $Sp\left( 2+2n,\mathds{R}\right) $. The unique
algebraically independent invariant polynomial in the $\mathbf{1+n}$
of $\U\left( 1,n\right) $ is quadratic:
\begin{equation}
\mathcal{I}_{2}=\frac{1}{2}\left[ q_{0}^{2}-q_{i}^{2}+\left(
p^{0}\right)
^{2}-\left( p^{i}\right) ^{2}\right] =\left| Z\right| ^{2}-Z_{i}\overline{Z}%
^{i}.  \label{I2-N=2}
\end{equation}

The general analysis of the Attractor Equations, BH charge orbits,
attractor \textit{moduli spaces} and split attractor of such a
theory has been performed in \cite
{Bellucci:2006xz,Ferrara:2007tu,Ferrara:2008ap,Ferrara:2010cw}; here
we recall it briefly, and further consider the ``small'' charge
orbit of such models.

\begin{enumerate}
\item  the ``large'' (rank-$2$) BPS charge orbit reads \cite{Bellucci:2006xz}
\begin{equation}
\mathcal{O}_{BPS,rank-2}=\frac{\U(1,n)}{\U(n)},~\text{dim}_{\mathds{R}}=2n+1,~%
\mathcal{I}_{2}>0.
\end{equation}
Thus, as for all ``large'' BPS charge orbits \cite{Ferrara:1997tw},
there is no associated attractor \textit{moduli space} or,
equivalently, the number of ``non-flat'' scalar directions along the
flow is $\#=2n$.

\item  the ``large'' (rank-$2$) non-BPS charge orbit (with $Z_{H}=0$) reads
\cite{Bellucci:2006xz}
\begin{equation}
\mathcal{O}_{nBPS,rank-2}=\frac{\U(1,n)}{\U(1,n-1)},~\text{dim}_{\mathds{R}%
}=2n+1,~\mathcal{I}_{2}<0.
\end{equation}
Thus, the associated attractor \textit{moduli space} reads
\begin{equation}
\mathcal{M}_{nBPS,rank-2}=\mathds{CP}^{n-1},~\#=2.\label{CERN-1}
\end{equation}

\item  the unique ``small'' (rank-$1$) BPS charge orbit reads
\begin{equation}
\mathcal{O}_{BPS,rank-1}=\frac{\U(1,n)}{\U(n-1)\times \U\left( 1\right)
\ltimes
\mathds{C}_{n}^{n-1}},~\text{dim}_{\mathds{R}}=2n+1,~\mathcal{I}_{2}=0,
\end{equation}
where the subscript denotes charge with respect to the $\U\left(
1\right) $
commuting factor of the stabilizer. Thus, the associated attractor \textit{%
moduli space} reads
\begin{equation}
\mathcal{M}_{BPS,rank-1}=\mathds{C}^{n-1},~\#=2.\label{ccern-1}
\end{equation}
\end{enumerate}

It is worth of notice that (non-compact forms of) $\mathds{CP}^{n}$
spaces as moduli spaces of string compactifications have appeared in
the literature, either as particular subspaces of complex structure
deformations of certain Calabi-Yau manifold
\cite{Ceresole:1993nz,Dixon:1989fj} or as moduli spaces of some
asymmetric orbifolds of Type II superstrings
\cite{Ferrara:1989nm,Ferrara:1989br,Dabholkar:1998kv,Kounnas:1997hi}, or of
orientifolds \cite{Frey:2002hf}.

\subsection{\label{N=3,d=4}$\mathcal{N}=3$}\label{sec:N3}

The (K\"{a}hler) scalar manifold is \cite{Castellani:1985ka}
\begin{equation}
\mathcal{M}_{\mathcal{N}=3,n}=\frac{G_{\mathcal{N}=3,n}}{H_{\mathcal{N}=3,n}}%
=\frac{\U\left( 3,n\right) }{\SU\left( 3\right) \times \U\left(
n\right) \times \U\left( 1\right) },~\text{dim}_{\mathds{R}}=6n.
\end{equation}
This theory cannot be uplifted to $D=5$, and it does not enjoy an
interpretation in terms of Jordan algebras.

The $3+n$ vector field strengths and their duals, as well as their
asymptotical fluxes, sit in the \textit{fundamental} $\mathbf{3+n}$
\ representation of the $\U$-duality group
$G_{\mathcal{N}=3,n}=\U\left(
3,n\right) $, in turn embedded in the symplectic group $Sp\left( 6+2n,%
\mathds{R}\right) $. The unique algebraically independent invariant
polynomial in the $\mathbf{3+n}$ of $\U\left( 3,n\right) $ is
quadratic, and it reads ($A=1,2,3$, $I=1,...,n$)
\cite{Ferrara:2008ap}:
\begin{equation}
\mathcal{I}_{2}=\frac{1}{2}\left[ q_{A}^{2}-q_{i}^{2}+\left(
p^{A}\right)
^{2}-\left( p^{i}\right) ^{2}\right] =\frac{1}{2}Z_{AB}\overline{Z}%
^{AB}-Z_{I}\overline{Z}^{I},  \label{tired-7}
\end{equation}
The general analysis of the Attractor Equations, BH charge orbits,
attractor \textit{moduli spaces} and split attractor of such a
theory has been performed in
\cite{Ferrara:2007tu,Ferrara:2008ap,Ferrara:2010cw}; here we recall
it briefly, and further consider the ``small'' charge orbit of this
theory (the results are also consistent with the $D=3$ analysis of
\cite {Bossard:2009at}).

\begin{enumerate}
\item  the ``large'' (rank-$2$) $\frac{1}{3}$-BPS charge orbit reads \cite
{Bellucci:2007ds}
\begin{equation}
\mathcal{O}_{\frac{1}{3}-BPS,rank-2}=\frac{\U(3,n)}{\U(2,n)},~\text{dim}_{%
\mathds{R}}=2n+5,~\mathcal{I}_{2}>0.
\end{equation}
The associated attractor \textit{moduli space}, as all the $\frac{1}{%
\mathcal{N}}$-BPS attractor moduli spaces of $\mathcal{N}\geqslant 3$%
-extended, $D=4$ supergravity theories \cite{Andrianopoli:1997pn},
is a quaternionic symmetric space (recall Eq. (\ref{CERN-1})):
\begin{equation}
\mathcal{M}_{\frac{1}{3}-BPS,rank-2}=\frac{\SU(2,n)}{\SU(2)\times
\SU(n)\times
\U(1)}=c\left( \mathds{CP}^{n-1}\right) =c\left( \mathcal{M}_{\mathcal{N}%
=2,mc,nBPS,rank-2}\right) ,~\#=2n,
\end{equation}
where ``$c$'' denotes the $c$-map \cite{Cecotti:1988qn}.

\item  the ``large'' (rank-$2$) non-BPS charge orbit (with $Z_{AB,H}=0$)
reads \cite{Bellucci:2007ds}
\begin{equation}
\mathcal{O}_{nBPS,rank-2}=\frac{\U(3,n)}{\U(3,n-1)},~\text{dim}_{\mathds{R}%
}=2n+5,~\mathcal{I}_{2}<0.
\end{equation}
Thus, the associated attractor \textit{moduli space} reads
\begin{equation}
\mathcal{M}_{nBPS,rank-2}=\frac{\U\left( 3,n-1\right) }{\SU\left(
3\right)
\times \U\left( n-1\right) \times \U\left( 1\right) }=\mathcal{M}_{\mathcal{N}%
=3,n-1},~\#=6.
\end{equation}

\item  the unique ``small'' (rank-$1$) $\frac{2}{3}$-BPS charge orbit reads
\begin{equation}
\mathcal{O}_{\frac{2}{3}-BPS,rank-1}=\frac{\U(3,n)}{\U(2,n-1)\times
\U\left(
1\right) \ltimes \mathds{C}_{n+2}^{2,n-1}},~\text{dim}_{\mathds{R}}=2n+5,~%
\mathcal{I}_{2}=0,
\end{equation}
where the subscript denotes charge with respect to the $\U\left(
1\right) $
commuting factor of the stabilizer. Thus, the associated attractor \textit{%
moduli space} reads (recall Eq. (\ref{CERN-1}))
\begin{eqnarray}
\mathcal{M}_{\frac{2}{3}-BPS,rank-1} &=&\frac{\SU(2,n-1)}{\SU(2)\times
\SU(n-1)\times \U(1)}  \notag \\
&=&c\left( \mathds{CP}^{n-2}\right) =c\left( \left. \mathcal{M}_{\mathcal{N}%
=2,mc,nBPS,rank-2}\right| _{n\rightarrow n-1}\right) ,~\#=2.
\label{ccern-2}
\end{eqnarray}
\end{enumerate}

\subsection{\label{N=5,d=4}$\mathcal{N}=5$}\label{sec:N5}

The (special K\"{a}hler) \textit{scalar manifold} is
\cite{deWit:1981yv}
\begin{equation}
\mathcal{M}_{\mathcal{N}=5}=\frac{G_{\mathcal{N}=5}}{H_{\mathcal{N}=5}}=%
\frac{\SU\left( 1,5\right) }{\SU\left( 5\right) \times \U\left( 1\right) },%
\text{~dim}_{\mathds{R}}=10.
\end{equation}
\textit{No} matter coupling is allowed (\textit{pure} supergravity).
This theory cannot be uplifted to $D=5$, but it is associated to the
Jordan triple system $%
M_{2,1}\left( \mathds{O}\right) $ generated by the $2\times 1$
vectors over $\mathds{O}$ \cite {Gunaydin:1983rk,Gunaydin:1983bi}.

The $10$ vector field strengths and their duals, as well as their
asymptotical fluxes, sit in the \textit{three-fold antisymmetric} irrepr. $%
\mathbf{20}$ of the $\U$-duality group $G_{\mathcal{N}=5}=\SU\left(
1,5\right) $. As discussed in \cite{Ferrara:2008ap}, unique
algebraically independent invariant polynomial in the $\mathbf{20}$
of $\SU\left( 1,5\right) $ is quartic in the bare charges (see
\textit{e.g.} the treatment of \cite {Ferrara:2008ap}), but is a
\textit{perfect square} of a quadratic
expression when written in terms of the scalar-dependent \textit{%
skew-eigenvalues }$\mathcal{Z}_{1}$ and $\mathcal{Z}_{2}$ of the
central charge matrix $Z_{AB}$ ($A=1,...,5$):
\begin{equation}
\mathcal{I}_{4}\left( p,q\right) \equiv Z_{AB}\overline{Z}^{BC}Z_{CD}%
\overline{Z}^{DA}-\frac{1}{4}\left( Z_{AB}\overline{Z}^{AB}\right)
^{2}=\left( \mathcal{Z}_{1}^{2}-\mathcal{Z}_{2}^{2}\right) ^{2}.
\label{ostia-2}
\end{equation}
This property distinguishes the $\mathcal{N}=5$ ``pure'' theory from
the previously treated $\mathcal{N}=2$, $D=4$ magic Maxwell-Einstein
theory associated to $\J_{3}^{\mathds{C}}$, whose $\U$-duality group
$\SU(3,3)$ is a different non-compact from of $\SU(6)$, and makes the
discussion of charge orbits much simpler.

The general analysis of the Attractor Equations, BH charge orbits
and attractor \textit{moduli spaces }of such a theory has been
performed in \cite {Andrianopoli:2006ub,Ferrara:2008ap}; here we
recall it briefly, and further consider the ``small'' charge orbit
of this theory (the results are also consistent with the $D=3$
analysis of \cite{Bossard:2009at}).

\begin{enumerate}
\item  the ``large'' (rank-$2$) $\frac{1}{5}$-BPS charge orbit reads \cite
{Bellucci:2007ds}
\begin{equation}
\mathcal{O}_{\frac{1}{5}-BPS,rank-2}=\frac{\SU(1,5)}{\SU(3)\times
\SU\left( 2,1\right)
},~\text{dim}_{\mathds{R}}=19,~\mathcal{I}_{4}>0.
\end{equation}
The associated attractor \textit{moduli space}, as all the $\frac{1}{%
\mathcal{N}}$-BPS attractor moduli spaces of $\mathcal{N}\geqslant 3$%
-extended, $D=4$ supergravity theories \cite{Andrianopoli:1997pn},
is a quaternionic symmetric space, namely the \textit{universal
hypermultiplet} space:
\begin{equation}
\mathcal{M}_{\frac{1}{5}-BPS,rank-2}=\frac{\SU(2,1)}{\SU(2)\times \U(1)}=%
\mathds{CP}^{2},~\#=6.
\end{equation}

\item  the unique ``small'' (rank-$1$) $\frac{2}{5}$-BPS charge orbit reads
\begin{equation}
\mathcal{O}_{\frac{2}{5}-BPS,rank-1}=\frac{\SU(1,5)}{\SU(3)\ltimes \mathds{R}%
^{8}},~\text{dim}_{\mathds{R}}=19,~\mathcal{I}_{4}=0\Leftrightarrow \mathcal{%
Z}_{1}=\mathcal{Z}_{2}.
\end{equation}
Thus, the associated attractor \textit{moduli space} reads
\begin{equation}
\mathcal{M}_{\frac{2}{5}-BPS,rank-1}=\mathds{R}^{8},~\#=2.\label{ccern-3}
\end{equation}
Note that the stabilizer of $\mathcal{O}_{\frac{2}{5}-BPS,rank-1}$
is the
same as the stabilizer of the rank-$3$ $\frac{1}{2}$-BPS orbit of the $%
\mathcal{N}=2$ magic theory associated to $\J_{3}^{\mathds{C}}$.
\end{enumerate}

By comparing Eqs. (\ref{ccern-1}), (\ref{ccern-2}) and
(\ref{ccern-3}), it follows that the $\mathcal{N}=2$
\textit{minimally coupled}, $\mathcal{N}=3$ matter-coupled and
$\mathcal{N}=5$ ``pure'' theories, besides the fact that they cannot
be uplifted to $D=5$, all share the property that the number of
``non-flat'' directions supported by the unique rank-$1$ charge
orbit is $2$.

\section*{Acknowledgments}

We would like to thank Duminda Dahanayake for useful discussions. The work of LB and SF is supported by the ERC Advanced Grant no. 226455 \textit{%
SUPERFIELDS}. Furthermore, the work of SF is also supported in
part by DOE Grant DE-FG03-91ER40662. The work of MJD is
supported by the STFC under rolling grant ST/G000743/1. LB is grateful for hospitality at the Theoretical Physics group at Imperial College London  and the CERN theory division  (where he was supported by the
above ERC Advanced Grant).


\appendix

\section{Orbit Stabilizers}\label{sec:app1}
In order to determine the stabilizers of the orbits we will use the
infinitesimal Lie action of $\Aut(\FTS)\cong \text{Conf}\left(
\frak{J}\right)$ acting on the corresponding representative
canonical forms. Hence, one needs to define the action of the Lie
algebra $\mathfrak{Aut}(\FTS(\J))$ in the $\Str_0(\J)$-covariant
basis. To this end, one can introduce the \emph{Freudenthal
product}, $\wedge:\FTS\times\FTS\rightarrow\Hom_\R(\FTS)$, which for
$x=(\alpha, \beta, A, B),\; y=(\delta,\gamma,C,D)$ is defined by \be
x\wedge y=\Phi(\phi, X, Y, \nu), \quad\textrm{where}\quad \left\{
\begin{array}{lll}
\phi    &=&-(A\vee D+B\vee C)\\
X           &=&-\half(B\times D-\alpha C-\delta A)\\
Y           &=&\half(A\times C-\beta D-\gamma B)\\
\nu     &=&\frac{1}{4}(\Tr(A,D)+\Tr(C,B)-3(\alpha\gamma+\beta\delta))\\
\end{array}\right.
\ee
and $A\vee B\in\Rstr{\J}$ is defined by $(A\vee B)C=\half\Tr(B,C)A+\frac{1}{6}\Tr(A,B)C-\half B\times(A\times C)$.  The action of $\Phi:\FTS  \rightarrow\FTS$ is given by
\be\label{eq:ftslieaction}
\Phi(\phi,X,Y,\nu)\begin{pmatrix}\alpha&A\\B&\beta\end{pmatrix}=\begin{pmatrix}\alpha\nu+(Y,B)&\phi A-\frac{1}{3}\nu A+2Y\times B +\beta X \\-^t\phi B+\frac{1}{3}\nu B+2X\times A+\alpha Y&-\beta\nu+(X,A)\end{pmatrix}.
\ee
The maps $\Phi\in\Hom_\R(\FTS)$ are in fact Lie algebra elements. Moreover, every Lie algebra element is given by some $\Phi$. More precisely we have the following theorem \cite{Yokota:2009}.
\begin{theorem}[Imai and Yokota, 1980]
\be
\mathfrak{Aut}(\mathfrak{F})=\{\Phi(\phi,X,Y,\nu)\in \Hom_{\R}(\mathfrak{F})|\phi\in\mathfrak{Str}_{0}(\J), X,Y\in\J, \nu\in\R\}.
\ee
where the Lie bracket
\be
[\Phi(\phi_1,X_1,Y_1,\nu_1),\Phi(\phi_2,X_2,Y_2,\nu_2)]=\Phi(\phi,X,Y,\nu)
\ee
is given by
\be
\begin{split}
\phi&=[\phi_1,\phi_2]+2(X_1\vee Y_2-X_2\vee Y_1)\\
X&=(\phi_1+\frac{2}{3}\nu_1)X_2-(\phi_2+\frac{2}{3}\nu_2)X_1\\
Y&=(\phi_2+\frac{2}{3}\nu_2)Y_1-(^t\phi_1+\frac{2}{3}\nu_1)Y_2\\
\nu&=\Tr(X_1,Y_2)-\Tr(Y_1,X_2).\\
\end{split}
\ee
\end{theorem}
We will frequently consider (see also \cite{Borsten:2011nq}) the Lie
algebra elements of the form \be \widehat{\Phi}(X, Y):=\Phi(0, X, Y,
0). \ee The Hermitian conjugate is defined by \be
\widehat{\Phi}^\dagger(X, Y)=\widehat{\Phi}(Y, X). \ee Hermitian
(resp. anti-Hermitian) generators are non-compact (resp. compact)
\cite{Bellucci:2006xz}.

\subsection{An Example : The Exceptional Magic Theory\label{sec:orbitsmagicFTS}}

As an example, which may be quite simply generalised to all models treated here, we examine the case of $\FTS(\JO)$. In order to
determine the stabilizers of the the orbits, we will use the
infinitesimal Lie algebra action \eqref{eq:ftslieaction} to fix the
Lie sub-algebras annihilating the the canonical forms presented in
\autoref{thm:Shuk} \cite{Shukuzawa:2006}. Note, in this specific
case the construction of the Lie algebra elements $\Phi(\phi, X, Y,
\nu)$ corresponds to the decomposition, \be
\begin{split}
E_{7(-25)}&\supset E_{6(-26)}\\
\rep{133}&\rightarrow \rep{1+27+27'+78}
\end{split}
\ee
where $\phi, X, Y$, and $\nu$ sit in  the  \rep{78, 27, 27'} and \rep{1}, respectively.

For all canonical forms  one obtains
\be
\Phi(x_{\textrm{can}})=\pmtwo{\nu}{\phi A_\text{can}-\frac{1}{3}\nu A_\text{can}}{X\times A_\text{can}+Y}{\Tr(Y, A_\text{can})},\quad\text{where}\quad x_\text{can}=\pmtwo{1}{A_\text{can}}{0}{0},
\ee
so  we may set the dilatation generator $\nu$ to zero throughout.

\paragraph{Rank 1:} $A_\text{can}=0$
\be
\Phi(x_1)=\pmtwo{0}{0}{Y}{0}
\ee
$\Rightarrow Y=0$ while $X$ and $\phi$ are unconstrained. Hence, the stability group is
\be
H_1=E_{6(-26)}\ltimes \R^{27},
\ee
where $E_{6(-26)}$ is generated by $\phi$ and the 27 translations are generated by $X$.
\paragraph{Rank 2a:} $A_\text{can}=(1,0,0)$
\be \Phi(x_{2a})=\pmtwo{0}{\phi A_\text{can}}{X\times
A_\text{can}+Y}{\Tr(Y, A_\text{can})} \ee From the $D=5$ analysis
\cite{Shukuzawa:2006} we know that the Lie sub-algebra of
$\Rstr{\JO}$ satisfying $\phi A_\text{can}=0$ has 36 compact, 9
non-compact semi-simple generators and 16 translational generators
giving $\mathfrak{so}(1,9)\oplus\R^{16}$. For the remaining $27+27$
generators we obtain the following constraints:
\begin{enumerate}
\item
\be
\Tr(Y, A_\text{can})=0 \Rightarrow y_{11}=0.
\ee
\item
\be
X\times A_\text{can} +Y=0\Rightarrow \pmthree{0}{0}{0}{0}{x_{33}}{-x_{23}}{0}{-\overline{x}_{23}}{x_{22}}=\pmthree{0}{-y_{12}}{-\overline{y}_{13}}{-\overline{y}_{12}}{-y_{22}}{-y_{23}}{-y_{13}}{-\overline{y}_{23}}{-y_{33}}
\ee
\end{enumerate}
This gives 1 compact and 9 non-compact semi-simple generators \be
\widehat{\Phi}(\tilde{X},\tilde{Y}), \ee where, writing $x_{22}=x+y$
and $x_{33}=x-y$, \be
\tilde{X}=\pmthree{0}{0}{0}{0}{x+y}{x_{23}}{0}{\overline{x}_{23}}{x-y},\quad
\tilde{Y}=\pmthree{0}{0}{0}{0}{-x+y}{x_{23}}{0}{\overline{x}_{23}}{-x-y}.
\ee These, together with the $36$ compact and $9$ non-compact
generators from $\mathfrak{so}(1,9)\subset\mathfrak{Str}_{0}(\JO)$,
give a total of 37 compact generators and 18 non-compact semi-simple
generators producing $\mathfrak{so}(2,9)$, where we have used the
fact that $\SO(m,n)$ has $[m(m-1)+n(n-1)]/2$ compact and $mn$
non-compact generators.

The other $1+16$ components of $X$ generate translations,
\be
X^{'}=\pmthree{x_{11}}{0}{0}{0}{0}{0}{0}{0}{0},\quad X^{''}=\pmthree{0}{x_{12}}{\overline{x}_{13}}{\overline{x}_{12}}{0}{0}{x_{13}}{0}{0},
\ee
where $X^{'}$ commutes with $\mathfrak{so}(2,9)$. The remaining $16+16$ translational generators transform as the spinor of $\mathfrak{so}(2,9)$.
Hence, the stability group is
\be
H_{2a}=\SO(2,9)\ltimes \R^{32}\times\R.
\ee

\paragraph{Rank 2b:} $A_\text{can}=(-1,0,0)$
\be
\Phi(x_1)=\pmtwo{0}{\phi A_\text{can}}{X\times A_\text{can}-Y}{\Tr(Y, A_\text{can})}
\ee
The analysis goes through as above but with the sign of $\tilde{Y}$ flipped. This gives a total of 45 compact and 10 non-compact semi-simple generators giving $\mathfrak{so}(1,10)$.
Hence, the stability group is
\be
H_{2b}=\SO(1,10)\ltimes \R^{32}\times\R.
\ee

\paragraph{Rank 3a:} $A_\text{can}=(1,1,0)$
\be \Phi(x_{3a})=\pmtwo{0}{\phi A_\text{can}}{X\times
A_\text{can}+Y}{\Tr(Y, A_\text{can})} \ee From the $D=5$ analysis
\cite{Shukuzawa:2006}, we know that the Lie sub-algebra of
$\Rstr{\JO}$ satisfying $\phi A_\text{can}=0$ has 36 compact
semi-simple generators and 16 translational generators, yielding
$\mathfrak{so}(9)\oplus\R^{16}$. For the remaining $27+27$
generators, we obtain the following constraints:
\begin{enumerate}
\item
\be
\Tr(Y, A_\text{can})=0 \Rightarrow y_{11}=-y_{22}.
\ee
\item
\be
\begin{split}
X\times A_\text{can} +Y=0&\Rightarrow \pmthree{x_{33}}{0}{-\overline{x}_{13}}{0}{x_{33}}{-x_{23}}{-x_{13}}{-\overline{x}_{23}}{x_{11}+x_{22}}=\pmthree{-y_{11}}{-y_{12}}{-\overline{y}_{13}}{-\overline{y}_{12}}{y_{11}}{-y_{23}}{-y_{13}}{-\overline{y}_{23}}{-y_{33}} \\
&\Rightarrow x_{33}=y_{11}=0.
\end{split}
\ee
\end{enumerate}
This gives 16 non-compact semi-simple generators, \be
\widehat{\Phi}(\tilde{X},\tilde{Y}), \ee where, \be
\tilde{X}=\tilde{Y}=\pmthree{0}{0}{\overline{x}_{13}}{0}{0}{x_{23}}{x_{13}}{\overline{x}_{23}}{0}.
\ee These, together with the $36$ semi-simple generators from
$\mathfrak{so}(9)\subset\mathfrak{Str}_{0}(\JO)$, give a total of 36
compact generators and 16 non-compact generators producing
$F_{4(-20)}$, which is a non-compact form of $\Aut(\JO)$.

The remaining $10$ components of $X$ generate translations which, together with the 16 preserved translational generators of $\Rstr{\JO}$, transform as the fundamental $\rep{26}$ of $F_{4(-20)}$.

Hence, the stability group is
\be
H_{3a}=F_{4(-20)}\ltimes \R^{26}.
\ee

\paragraph{Rank 3b:} $A_\text{can}=(-1,-1,0)$
\be \Phi(R_1)=\pmtwo{0}{\phi A_\text{can}}{X\times
A_\text{can}-Y}{\Tr(Y, A_\text{can})} \ee The analysis goes through
as above, but with the sign of $\tilde{Y}$ flipped so that the 16
previously non-compact semi-simple generators become compact giving
the compact form $F_{4(-52)}=\Aut(\JO)$. Hence, the stability group
is \be H_{3a}=F_{4(-52)}\ltimes \R^{26}. \ee

\paragraph{Rank 4a:} $A_\text{can}=(-1,-1,-1)$
\be \Phi(x_{4a})=\pmtwo{0}{\phi A_\text{can}}{X\times
A_\text{can}+Y}{\Tr(Y, A_\text{can})} \ee From the $D=5$
analysis we know that the Lie sub-algebra of
$\Rstr{\JO}$ satisfying $\phi A_\text{can}=0$ has 52 compact
semi-simple generators giving $F_{4(-52)}$. For the remaining
$27+27$ generators, we obtain the following constraints:
\begin{enumerate}
\item
\be
\Tr(Y, A_\text{can})=0 \Rightarrow y_{11}+y_{22}+y_{33}=0.
\ee
\item
\be\label{eq:r41}
X\times A_\text{can} +Y=0\Rightarrow \pmthree{x_{11}}{x_{12}}{\overline{x}_{13}}{\overline{x}_{12}}{x_{22}}{x_{23}}{x_{13}}{\overline{x}_{23}}{-(x_{11}+x_{22})}=\pmthree{-y_{11}}{-y_{12}}{-\overline{y}_{13}}{-\overline{y}_{12}}{-y_{22}}{-y_{23}}{-y_{13}}{-\overline{y}_{23}}{(y_{11}+y_{22})},
\ee
where we have abused the notation by use the same symbols for $X,Y$ after imposing the condition $\Tr(Y)=0$. We have also used the identity $X\times (-\mathds{1})=X-\Tr(X)\mathds{1}$  so that $X\times A_\text{can} +Y=0$ implies $\Tr(X)=0$, therefore giving the implication in \eqref{eq:r41}.
\end{enumerate}
This gives 26 compact semi-simple generators, \be
\widehat{\Phi}(\tilde{X}, \tilde{Y}), \ee where \be
\tilde{X}=\pmthree{x_{11}}{x_{12}}{\overline{x}_{13}}{\overline{x}_{12}}{x_{22}}{x_{23}}{x_{13}}{\overline{x}_{23}}{-(x_{11}+x_{22})},\quad
\tilde{Y}=
\pmthree{-x_{11}}{-x_{12}}{-\overline{x}_{13}}{-\overline{x}_{12}}{-x_{22}}{-x_{23}}{-x_{13}}{-\overline{x}_{23}}{(x_{11}+x_{22})}.
\ee These, together with the $52$ compact  semi-simple generators
from $F_{4(-52)}$, give a total of 78 compact generators producing
$E_{6(-78)}$.

Hence, the stability group is
\be
H_{4a}=E_{6(-78)}.
\ee

\paragraph{Rank 4b:} $A_\text{can}=(1,1,-1)$
\be \Phi(x_{4b})=\pmtwo{0}{\phi A_\text{can}}{X\times
A_\text{can}+Y}{\Tr(Y, A_\text{can})} \ee From the $D=5$ analysis
\cite{Shukuzawa:2006}, we know that the Lie sub-algebra of
$\Rstr{\JO}$ satisfying $\phi A_\text{can}=0$ has 36 compact and 16
non-compact semi-simple generators giving $F_{4(-20)}$. For the
remaining $27+27$ generators, we obtain the following constraints:
\begin{enumerate}
\item
\be
\Tr(Y, A_\text{can})=0 \Rightarrow y_{11}+y_{22}=y_{33}.
\ee
\item
\be
X\times A_\text{can} +Y=0\Rightarrow \pmthree{x_{11}}{x_{12}}{-\overline{x}_{13}}{\overline{x}_{12}}{x_{22}}{-x_{23}}{-x_{13}}{-\overline{x}_{23}}{x_{11}+x_{22}}=\pmthree{-y_{11}}{-y_{12}}{-\overline{y}_{13}}{-\overline{y}_{12}}{-y_{22}}{-y_{23}}{-y_{13}}{-\overline{y}_{23}}{-(y_{11}+y_{22})}.
\ee
\end{enumerate}
This gives 10 compact and 16 non-compact semi-simple generators, \be
\widehat{\Phi}(\tilde{X}, \tilde{Y}), \ee where \be
\tilde{X}=\pmthree{x_{11}}{x_{12}}{\overline{x}_{13}}{\overline{x}_{12}}{x_{22}}{x_{23}}{x_{13}}{\overline{x}_{23}}{x_{11}+x_{22}},\quad
\tilde{Y}=
\pmthree{-x_{11}}{-x_{12}}{\overline{x}_{13}}{-\overline{x}_{12}}{-x_{22}}{x_{23}}{x_{13}}{\overline{x}_{23}}{-(x_{11}+x_{22})}.
\ee These, together with the $36$ compact and 16 non-compact
semi-simple generators from $F_{4(-20)}$, give a total of 46 compact
generators and 32 non-compact generators producing $E_{6(-14)}$.

Hence, the stability group is
\be
H_{4b}=E_{6(-14)}.
\ee

\paragraph{Rank 4c:} $A_\text{can}=(1,1,1)$
\be \Phi(x_{4c})=\pmtwo{0}{\phi A_\text{can}}{X\times
A_\text{can}+Y}{\Tr(Y, A_\text{can})} \ee From the $D=5$ analysis
\cite{Shukuzawa:2006}, we know that the Lie sub-algebra of
$\Rstr{\JO}$ satisfying $\phi A_\text{can}=0$ has 52 compact
semi-simple generators giving $F_{4(-52)}=\Aut(\JO)$. For the
remaining $27+27$ generators, we obtain the following constraints:
\begin{enumerate}
\item
\be
\Tr(Y, A_\text{can})=0 \Rightarrow y_{11}+y_{22}+y_{33}=0.
\ee
\item
\be
X\times A_\text{can} +Y=0\Rightarrow \pmthree{-x_{11}}{-x_{12}}{-\overline{x}_{13}}{-\overline{x}_{12}}{-x_{22}}{-x_{23}}{-x_{13}}{-\overline{x}_{23}}{x_{11}+x_{22}}=\pmthree{-y_{11}}{-y_{12}}{-\overline{y}_{13}}{-\overline{y}_{12}}{-y_{22}}{-y_{23}}{-y_{13}}{-\overline{y}_{23}}{y_{11}+y_{22}}.
\ee
\end{enumerate}
This gives 26 non-compact semi-simple generators, \be
\widehat{\Phi}(\tilde{X}, \tilde{Y}), \ee where \be
\tilde{X}=\tilde{Y}=
\pmthree{x_{11}}{x_{12}}{\overline{x}_{13}}{\overline{x}_{12}}{x_{22}}{x_{23}}{x_{13}}{\overline{x}_{23}}{-(x_{11}+x_{22})}.
\ee These, together with the $52$ compact semi-simple generators
from $F_{4(-52)}$, give a total of 52 compact generators and 26
non-compact generators producing $E_{6(-26)}=\Str_0(\JO)$.

Hence, the stability group is
\be
H_{4c}=E_{6(-26)}.
\ee

\bigskip

This procedure can be repeated for all magical theories, yielding
the results
reported in Table 6, as well as for all $\mathcal{N}=2$, $D=4$ \textit{%
symmetric }supergravity theories with a Jordan algebraic
interpretation (see also the treatment of \cite{Borsten:2011nq}). For
the $D=5$ treatment, see \cite{Shukuzawa:2006}.


%

\end{document}